\documentclass{article}
\pdfoutput=1 
\usepackage{amsmath, amsthm}
\usepackage[utf8]{inputenc}
\usepackage{graphicx}
\usepackage{nomencl}
\usepackage{acronym}
\usepackage{times}
\usepackage[T1]{fontenc}
\usepackage[english]{babel}
\usepackage{authblk}
\usepackage{tikz} 
\usepackage{xkeyval}
\usepackage{setspace}
\usepackage{color}
\usepackage{epstopdf}
\usepackage[hypertexnames=false]{hyperref}
\usepackage[all]{hypcap} 
\usepackage[citestyle=authoryear,natbib=true,backend=bibtex]{biblatex}
\usepackage{amssymb,epsfig}
\usepackage{graphicx}
\usepackage{titlesec}
\usepackage{authblk}

\providecommand{\keywords}[1]{\textbf{\textit{Keywords: }} #1}

\hypersetup{%
   pdftitle={Reconstructing eruptive source parameters from tephra deposit},
   pdfauthor={A. Spanu},
   pdfkeywords={Tephra deposit; eruptive parameters; numerical modelling; grain size distribution; granulometry; erupted mass.},
}%

\let\origappendix\appendix 
\renewcommand\appendix{\clearpage\pagenumbering{roman}\origappendix}

\bibliography{bibl.bib}

\makeatletter
\renewcommand\listoftables{%
    \section{\listtablename}%
      \@mkboth{%
          \MakeUppercase\listtablename}%
         {\MakeUppercase\listtablename}%
    \@starttoc{lot}%
    }
\makeatother

\title{Reconstructing eruptive source parameters from tephra deposit: a numerical approach for medium-sized explosive eruptions}

\author[1,2,+]{A. Spanu \footnote{Corresponding author: antonio.spanu@dlr.de}}
\author[1]{M. de' Michieli Vitturi}
\author[1,*]{S. Barsotti}
\affil[1]{Istituto Nazionale di Geofisica e Vulcanologia\\ Sezione di Pisa, Italy\\}
\affil[2]{ Scuola Normale Superiore di Pisa\\ Italy}
\affil[+]{Now at Deutschen Zentrums fur Luft- und Raumfahrt, Germany } 
\affil[*]{Now at Icelandic Meteorological Office, Iceland }

\begin{document}

\maketitle
\begin{refsection}
\begin{abstract}
Since the seventies, several reconstruction techniques have been proposed, and are currently used, to extrapolate and quantify eruptive parameters from sampled deposit datasets. Discrete numbers of tephra ground loadings or stratigraphic records are usually processed to estimate source eruptive values. Reconstruction techniques like Pyle, Power law and Weibull are adopted as standard to quantify the erupted mass (or volume) whereas Voronoi for reconstructing the granulometry. Reconstructed values can be affected by large uncertainty due to complexities occurring within the atmospheric dispersion and deposition of volcanic particles. Here we want to quantify the sensitivity of reconstruction techniques, and to quantify how much estimated values of mass and grain size differ from emitted and deposited ones. We adopted a numerical approach simulating with a dispersal code a mild explosive event occurring at Mt. Etna, with eruptive parameters similar to those estimated for eruptions occurred in the last decade. Then we created a synthetic deposit by integrating the mass on the ground computed by the model over the computational domain (>50000 km$^2$). Multiple samplings of the simulated deposit are used for generating a large dataset of sampling tests afterwards processed with standard reconstruction techniques. Results are then compared and evaluated through a statistical analysis, based on 2000 sampling tests of 100 samplings points. On average, all the used techniques underestimate deposited and emitted mass. A similar analysis, carried on Voronoi results, shows that information on the total grain size distribution is strongly deteriorated. 
In particular, particles smaller than $ \phi =3$ are strongly under-represented in the final total grain size distribution  In addition, when the proximal area is excluded from the computation (<4km from the vent), an important fraction of coarse particles is missed. 
Here we present a new method allowing an estimate of the deficiency in deposited mass for each simulated class. Finally a sensitivity study on eruptive parameters is presented in order to generalize our results to a wider range of eruptive conditions.
\end{abstract}

\keywords{Tephra deposit; eruptive parameters; numerical modelling; grain size distribution; granulometry; erupted mass.}

\section{Introduction}
During explosive volcanic eruptions, solid particles of various dimensions, ranging from several cm to a few microns (tephra) are injected into the atmosphere. Particle are subsequently
settled toward the ground under the action of atmospheric dynamics, aggregative processes, and gravity. 
Atmospheric transport process, depending on particle characteristics (size, density, sphericity) and altitude of release, can last from a few hours up to weeks. Area affected by ash fallout can extend thousands of square kilometers around the vent \citep{Sparks_et_al_1997}.
A primary objective of most volcanologists is to quantify and to measure the magnitude and the intensity of an eruption. 
Erupted mass is fundamental in order to deepen our comprehension of the mechanisms triggering a volcanic eruption, the column evolution, and the ash cloud dynamics.
During a volcanic crisis several direct measurements can provided information about the ongoing activity. 
Remote sensing instruments are currently adopted by volcanologists to give estimates of column height \citep{Rose_et_al_1995, Petersen_and_Arason_2010}, duration of the event \citep{Johnson_et_al_2004}, gas and particles exit velocity \citep{Dubosclard_et_al_2004}, and plume composition \citep{Rose_et_al_2000, Spinetti_et_al_2013}.  Some of these methods provide estimates with good resolution and details on the measured quantity.
Nevertheless, after an eruption, the deposit is still one of the main products to analyze in order to attempt a cumulative estimation of erupted mass and grain-size distribution. 
Historically, the integration of discrete samplings of deposit thickness values (or mass loading values) has been used to quantify the volume (or mass) of solid erupted material and, in this way, to assign a size and style to an eruption \citep{Fisher_1964, Walker_1973}. 
In past decades, several techniques have been proposed to optimize the integration of field data and to obtain an estimate of the emitted material (see for example \citet*{Pyle_1989}, and \citet*{Fierstein_and_Nathenson_1992}). 
More recently, \citet*{Bonadonna_and_Houghton_2005} and \citet*{Bonadonna_and_Costa_2012} introduced other methodologies for estimating volume and total grain size distribution, currently adopted by volcanologists during their field studies. 
Furthermore, \citet*{ Burden_et_al_2013, Engwell_et_al_2013} used statistical methods to study the uncertainty associated with tephra thickness measurements and volume estimation . 
Indeed, when estimating the amount of erupted material and grain-size distribution from discrete sampling the information that we can obtain directly from the tephra deposit is just a part of the total emitted by the source. 
Very few attempts have been made to interpret and revise the ground information in order to better constrain the initial eruptive conditions: \citet*{Connor_and_Connor_2006} used an inversion techniques and recently \citet*{Gudmundsson_et_al_2012, Stevenson_et_al_2015} performed an integration of ground measurements and satellite observations.
The importance of estimating eruptive source parameters (GSD and erupted mass) with a good accuracy is also due to the growing use of dispersal codes for ash hazards assessment purposes \citep{Sparks_et_al_1997, Textor_et_al_2005, Folch_2012, Fagents_et_al_2013}. Indeed, numerical models for ash dispersion are generally initialized with source conditions inferred from field observations. 
Numerical codes can also help in constraining eruptive source conditions by comparing numerical results with measured data. In  \citet*{Burden_et_al_2011} and \citet*{Biass_and_Bonadonna_2011}, the estimation of column height as a function of depositional clast size has been inferred by the use of a numerical code. 
\citet*{Klawonn_et_al_2012}, also inverted deposit data to better constrain eruptive parameters with a dispersal model. 
Quite recently,\citet*{Woodhouse_et_al_2013, Bursik_et_al_2012} investigated wind effect on column dynamics and height revealing the importance of its consideration to avoid strong under-estimation of mass flow-rate. 
Indeed, a numerical simulation has the advantage to keep all main initial conditions under control, and to follow the evolution of the main macroscopic features of investigated process. 
Here we use a computational approach to explore the quality of information contained in a tephra ground deposit and to test the efficacy of reconstruction techniques in quantifying eruptive parameters, used as code input. We pursue this objective by simulating a fictitious eruptive event at Mt. Etna for which initial source conditions are defined a priori and calculating its deposit as described in the Fig.  \ref{sketch}.
\begin{figure}[!htb]
\centering{\includegraphics[width=1.0\textwidth]{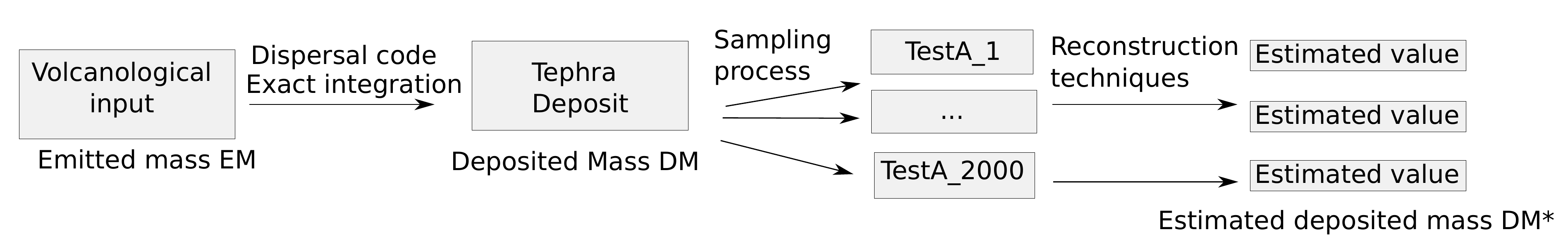}}
\caption{\label{sketch} Schematic representation of the computational strategy adopted.}
\end{figure}
Deposited masses for each grain size (DM$_i$) are computed and compared with the known corresponding masses emitted at the vent (EM$_i$). 
Furthermore, we randomly sample the deposit and apply several reconstruction techniques to obtain an estimate (DM$_i^*$) of the deposited mass. This is repeated multiple times, in order to perform a statistical analysis on the results.

\section{Adopted method}
\subsection{Numerical model}
To create a synthetic deposit we adopt the dispersal code VOL-CALPUFF \citep{Barsotti_et_al_2008a} simulating volcanic ash transport and deposition in a transient and 3D atmosphere. 
The model couples an Eulerian description of the initial plume rise phase, where plume theory equations are solved \citep{Bursik_2001}, with a Lagrangian description of the transport of material leaving the eruptive column. 
At each step particles are released from the column as a series of Gaussian packets (puffs) which are transported and diffused by the wind during their fall toward the ground due to gravity.
Tracking puff movements within the 3D domain, the code computes at each time step the amounts of mass advected out of the domain, still suspended in the atmosphere and deposited on the ground. VOL-CALPUFF model has been tested and adopted to simulate volcanic ash dispersal and deposition at several volcanoes worldwide \citep{Barsotti_and_Neri_2008b, Barsotti_et_al_2011, Spinetti_et_al_2013}, confirming how this model is a well-established tool for simulating eruptive events of variable sizes and occurring over a large range of spatial scales.

For the application shown in this paper, the VOL-CALPUFF model has been initialized with meteorological input data produced by the non-hydrostatic code LAMI \citep{Doms_and_Schattler_2002} with 7km horizontal resolution and 23 vertical levels and refined down to one km horizontally and one hour in time by the processor CALMET \citep{Scire_et_al_2000}.

\subsection{Recontruction techniques}
\subsubsection{Erupted mass}\label{Erupted mass}
The determination of the amount of erupted material is historically made starting from field measurements collected over the domain at discrete points \citep{Walker_1973, Pyle_1989, Fierstein_and_Nathenson_1992}. These data can be either measurement of deposit thickness for historical eruptions or of mass loading for more recent ones. 
Several methods have been proposed to estimate erupted volume from deposit data \citep{Rose_et_al_1973, Pyle_1989, Fierstein_and_Nathenson_1992, Sulpizio_2005, Bonadonna_and_Costa_2012, Burden_et_al_2013}. 
Estimation process consists on several steps, as suggested by the IAVCEI Commission on Tephra Hazard Modelling \citep{Bonadonna_et_al_2011a}.
First step is drawing a discrete number (N) of isopach (or isomass) contours, usually using an interpolation techniques.
Second step is calculating the squared roots of the obtained isopach areas ($x_i$) corresponding to $t_i$ thickness.
The main idea is to express $t_i$ as a function of the squared roots $x_i$ of its corresponding isopachs area. 
This is done by a fitting procedure, where the function $t(x)$ is chosen from a fixed family, described by one or more parameter, by minimizing a residual function.
Finally the eruptive volume is calculated by integrating the function $t(x)$ over the domain.
Recently an alternative to these steps, avoiding drawing isomass maps, has been proposed in \citet{Burden_et_al_2013} using a statistical method on the sampling points measurements.

Within the first category of methods, in this work we test the following families of fitting functions (see also Table 1):
\begin{table}
\caption{\label{Table1} Reconstruction methods presented with explicit formulas for mass loading (t) as a function of corresponding isomass areas (x), and residual cost functions. 
N is the number of isomass. For Weibull methods (referring to Weibull 1, 2 and 3), parameters have been chose, according with \citep{Bonadonna_and_Costa_2012}, in the best initial range $ \lambda \in [.1, 1000] km$, $ \theta \in [.1, 5000] cm$ and $ n \in [0.2, 2] $.}
\begin{center}\small{ \begin{tabular}{c c cc }
Method & Formula t(x) & Residual function & Free parameters\\
\hline\hline
Pyle & $ t_0 \exp(m x)$ & $ \sum_{i=1, N} (log(t(xi))-log(t_i))^2$ & 6 \\
Power law & $ t_p x_m $ & $ \sum_{i=1, N} (log(t(xi))-log(t_i))^2$ & 2 +2 arb\\
Weibull 1 & $ \theta(x/\lambda)^{n-2} \exp(-(x/\lambda)^n)$ & $ \sum_{ i=1, N} (t(x_i)-t_i)^2$ &3 \\
Weibull 2 & $ \theta(x/\lambda)^{n-2} \exp(-(x/\lambda)^n)$ & $ \sum _{i=1, N} 1/t_i^2 (t(x_i)-t_i)^2$ &3\\
Weibull 3 & $ \theta(x/\lambda)^{n-2} \exp(-(x/\lambda)^n) $ & $ \sum _{i=1, N}1/t_i (t(x_i)-t_i)^2 $ &3\\
\end{tabular}
}\end{center}
\end{table}

1) Exponential or Pyle’s method \citep{Pyle_1989}: This method is based on the assumption that the thickness, calculated on elliptical isopachs, follows an exponential decay with distance from the vent. 
In \citet{Fierstein_and_Nathenson_1992} and \citet{Bonadonna_et_al_1998} A modification of the method is proposed, mainly to account for the break-in-slope of some tephra deposits, by using various exponential segments. 
This method is now widely used even though it is very sensitive to the number and the choice of straight segments. 
Here we utilize three segments and an automatic procedure is defined to compute their bounds in order to minimize the residual function, thus reducing the arbitrariness of the choice.

2) Power law method \citep{Bonadonna_and_Houghton_2005}: In their paper the authors suggest a power-law best fit of field data to integrate the total erupted volume. 
Associated volume can be calculated as: $ V=\int 2 t_{pl} x^{m+1} dx$ , but this integral, when evaluated over the interval $ [0, +\infty]$ , is infinite. 
To avoid this problem such an interval is replaced by the arbitrary one $ [x_0, x_1]$. 
As already stated by \citet{Bonadonna_and_Costa_2013} Power Law results are strongly influenced by this choice.  
For this reason, in our statistical analysis average values of these distances are fixed at 1km and 300 km (accordingly with the eruption size) with a range of variability equal to $ \pm10\%$ .

3) Weibull method \citep{Bonadonna_and_Costa_2012, Bonadonna_and_Costa_2013}: Recently a new method based on Weibull functions has been proposed by Bonadonna and Costa. As main assumption the model describes the function $ xT(x)$ with a Weibull distribution and three main parameters ($\lambda , \theta, n $).
$ \lambda$  represents the characteristic decay length scale of deposit thinning, $\theta$ represents a thickness scale and $n$ is a shape parameter (dimensionless). 
In this work, model outcomes are expressed as ground loadings, so we apply all the previous methods for the mass estimation by considering isomass contours instead of isopachs as frequently done in case of fresh deposit \citep{Scollo_et_al_2007, Andronico_et_al_2008a, Andronico_et_al_2009}.
\subsubsection{Granulometry}\label{Granulometry}
Total grain-size distribution (TGSD) of tephra-fall deposits is a crucial eruptive parameter for tephra-dispersal modelling and hazard mitigation plans \citep{Scollo_et_al_2008, Folch_2012} and, several techniques have been developed to reconstruct it on the basis of ground information \citep{Murrow_et_al_1980, Bonadonna_and_Houghton_2005}. 
Usually, grain size are expressed in $\phi$ scale where $ \phi = - log_2(d/d_0)$ , d is the particle diameter and d$_0$  is a reference diameter of 1 mm.
After \citet{Bonadonna_and_Houghton_2005}, Voronoi technique has been adopted by the volcanological community as a benchmark for reconstructing TGSD. 
For this reason, it has been chosen in our work for a comparison with the exact emitted and deposited granulometry.
Voronoi, or nearest-neighbor, technique is a constant piecewise interpolation method, built upon the Voronoi tessellation. 
Given a set of points called seeds, the plane, over which they lay, is partitioned into convex cells, each one consisting of all those points closer to that seed than to any others. 
Then, as proposed in \citet{Bonadonna_and_Houghton_2005}, the grain size and mass load of each sample point is assigned to the corresponding cell and total grain-size distribution is obtained from the mass-weighted (where the mass is defined as the product between the cell area and the cell ground loading) average of all the sampled values over the whole deposit. 
The main requirement to apply Voronoi method is to fix the deposit extent, adding to the original data set also deposit zero values. 
This step is essential to prevent the external Voronoi cells from having unlimited area. Dependence of the reconstructed granulometry on the location of the zeros has been previously recognized and discussed in \citep{Bonadonna_and_Houghton_2005}.

\section{Application to Mt. Etna}
\begin{figure}[!htb]
\centering{\includegraphics[width=1.0\textwidth]{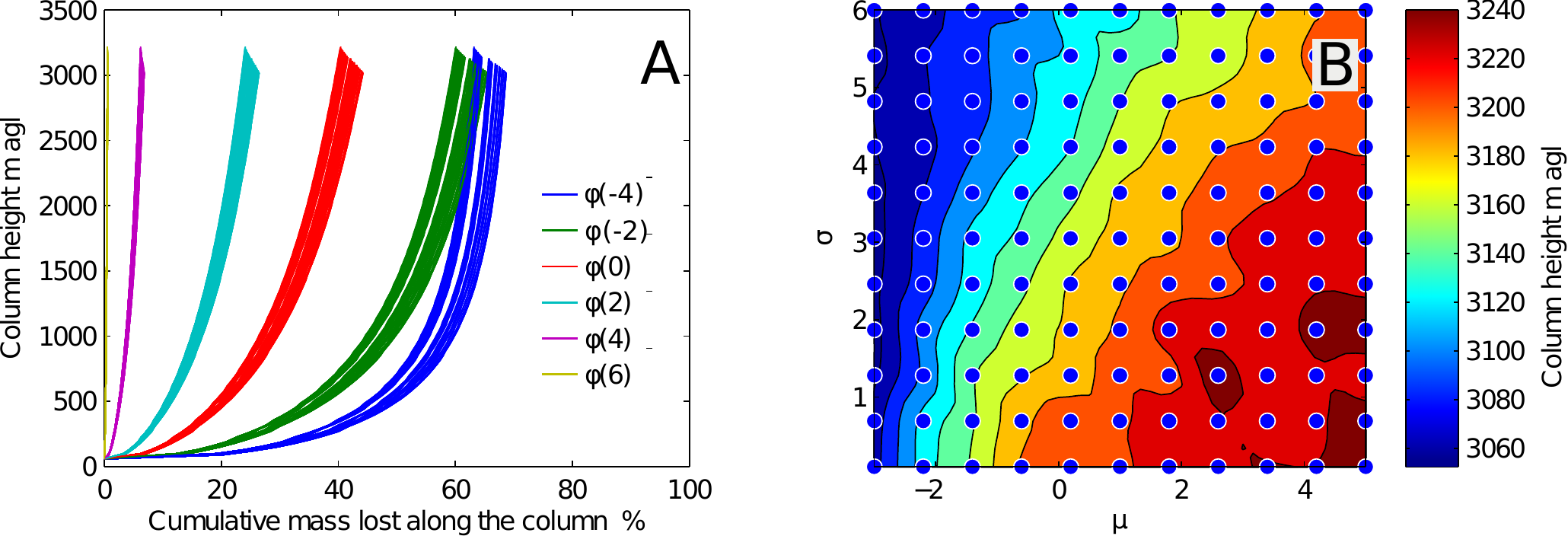}}
\caption{\label{Fig1} A) Vertical distribution of released mass along the column for different class sizes (blue, yellow, red and magenta) expressed as a percentage of the total (EM$_i$).
Lines are the result of different runs made by varying the parameters $\mu$ and $\sigma$ describing the initial GSD. B) Column height is expressed as responding function of initial sampled initial conditions.}
\end{figure}

Mt. Etna is the most active volcano in Europe, and each year it shows a wide variability in its eruptive activity. 
It has been extensively studied, and a wide literature exists regarding past activities, eruptive styles and products \citep{Branca_and_Del_Carlo_2005, Andronico_et_al_2005, Scollo_et_al_2007, Andronico_et_al_2008a, Andronico_et_al_2009, Andronico_et_al_2014}. 
It is definitely an open-pit laboratory of volcanology and for all these reasons we chose it as our study test case. 
In this paper, VOL-CALPUFF code is used to simulate a fictitious event representative of eruption conditions occurred on 24 November 2006 during which a six-hour-long eruption occurred \citep{Andronico_et_al_2009}.
At that time a quite persistent wind was blowing toward south-east with intensities up to $ 10 $ m/s at $ 3$ km above the vent. 
Due to these meteorological conditions, the investigated domain, here fixed to 272x277 km$^2$, is displaced south-east of the vent. 
All the main volcanological parameters, (e.g. mass eruption rate) are fixed a priori in order to simulate a mild eruption producing a 3-km high tilted column above the vent lasting for 6 hours. To this aim, the mass flow-rate has been fixed to $ 1\cdot 10^5$ kg/s,  with total emitted mass of $ 2.5\cdot 10^9$ kg. Simulated total grain size distribution is in $\phi$  range  $[$-5, 6$]$ ( 32 mm to  16$\mu$m). 
Resulting column height and ground loading values are consistent with observation \citep{Andronico_et_al_2008b} and more general with those observed for recent eruptive activities at Etna in: 2001 \citep{Scollo_et_al_2007}, 2002 \citep{Andronico_et_al_2008a}, 2006 \citep{Andronico_et_al_2009} .
Fig. \ref{Fig1}A shows, for different size classes, the vertical distribution of cumulative mass released along the column. Mass lost along the column is calculated for each class as a function of settling velocities.

Settling velocity is computed as a function of particle Reynolds number and depends on particle characteristics (dimension, density, sphericity) as well as atmospheric properties (air density and viscosity), according with the modification of \citet*{Wilson_and_Huang_1979} model as presented on \citet*{Pfeiffer_et_al_2005}.
Several other settling velocity models \citep{Pfeiffer_et_al_2005, Bonadonna_et_al_1998} have been tested showing similar results.

A sensitivity analysis has been carried out assuming a normal TGSD with parameters $\mu$ and $\sigma$ ranging in the intervals $[-3,5]$ and $[0.5,6]$ respectively. Initial distribution has been then partitioned in 24 classes equally spaced in the $\phi$ scale, while the other input parameters have been kept fixed as described before.

\begin{figure}[!htb]
\centering{\includegraphics[width=1.0\textwidth]{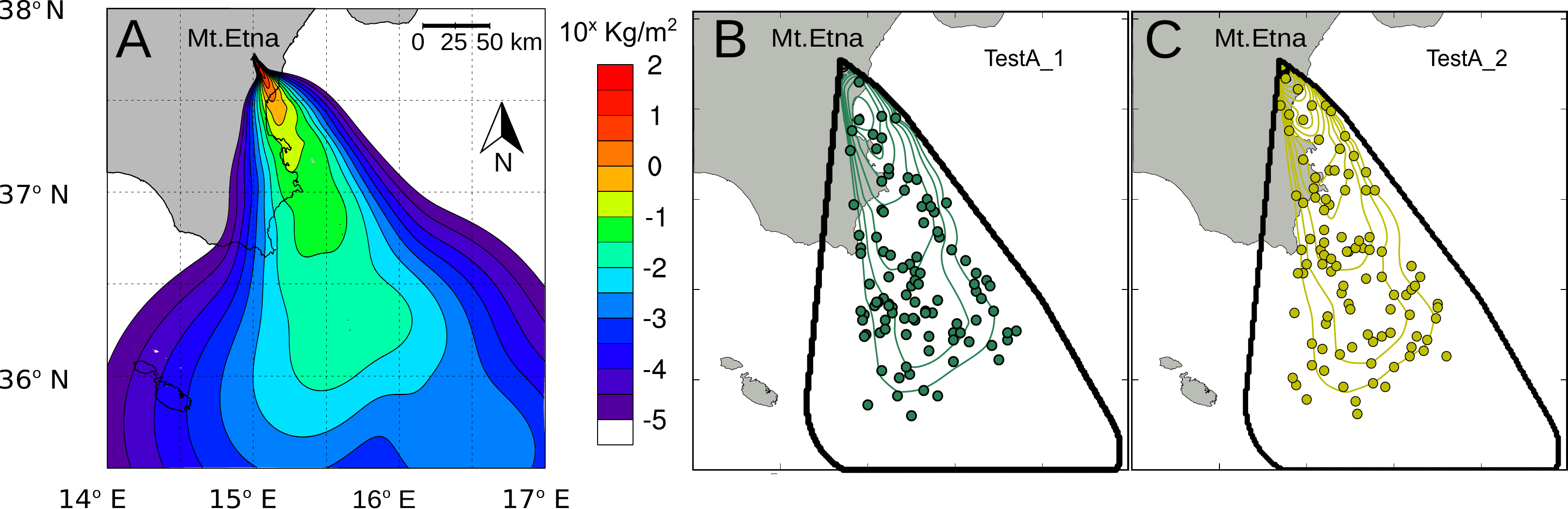}}
\caption{\label{Fig2} A) Computed ground loading contours [kg/m$^2$] for the synthetic deposit at the end of the simulation. 
Isomass contours are log-spaced from 10$^{-5}$ to 10$^{2}$ [kg/m$^2$]. B,C) Examples of sampling tests (colored circle) with the corresponding reconstructed deposit using NN.
Dark crosses represent the assumed zero values. The black triangle indicates the position of the eruptive vent.}
\end{figure}

Fig. \ref{Fig2}B presents the results of the sensitivity analysis, showing that column height is slightly sensitive to the size distribution. For this reason, without any loss in result generality, we assumed an initial granulometry with a uniform distribution. 
This choice does not affect results also because, assuming particles do not interact during the transport, for each class the percentage of deposited mass with respect to the emitted one is independent from the assumed initial granulometry (the independent deposits for each class are presented in the Auxiliary Material). 
As a consequence our results can be easily extended to different distributions of particles leaving the column. 

Dispersal model produces a numerical result on 250$\times$255 computational cells. 
At each cell center is associated a ground loading value (mass per unit area), obtained from mass integration of Gaussian packets deposited on the ground over the cell. 
This integration is performed by using a Taylor formula with a maximum numerical error fixed a priori (smaller than machine precision). In the following we will refer to this integration of Gaussian functions with the term “exact integration”. In this way we calculate, for each simulated particle size, the cumulative amount of mass deposited over the computational domain. 

Fig. \ref{Fig2}A shows the simulated tephra deposit calculated after 24 hours. This time frame is a reasonable time to allow the ash cloud leaving the computational domain and in real events to prevent further modification of the deposit by wind action. 
Only a small part of the deposit interests the inland while a large amount falls into the sea. 
Deposit cross-wind extension mainly depends on wind temporal stability and direction during the eruptive event. Due to wind variability on direction with altitude and with time the simulated eruption produces a wide deposit, which enlarges downwind the vent (Fig. \ref{Fig2}A).
In our case the column injects material up to $3$ km above the vent and the particles falling close to the vent produce ground loading values larger than $ 10^2 $ kg/m$^2$. 
Far away ground loading decreases but still shows values larger than $1$ kg/m$^2$ at about $30$ km downwind of the volcano (close to the main city of Catania and the International Airport “Fontanarossa”).

To test reconstruction techniques the whole set of 250$\times$255 integral loading values is sampled to create 2000 smaller data sets (“sampling tests”) of 100 points over which different techniques are applied.  
For each sampling test, points are generated using a cylindrical coordinate system and the sampling procedure is performed using a uniform probability distribution along the supposed major dispersal inferred main direction axis and a normal distribution for the angular coordinate. 
Major axis is calculated based on output atmospheric concentration and wind direction as usually done in sampling field work for recent eruptions \citep{Bonadonna_and_Houghton_2005}. Due to wind direction variability during the eruptive event, and therefore to the difficulty to find a major dispersal axis, a small random variability is considered on the angle in each sampling test. 
The area proximal to the vent is excluded by the sampling, but at least a point within 4 kilometers from the vent is considered in each test. We also sampled only inside the 10$ ^{-2}$ kg/m$^2$ isomass contour (corresponding to 0.01mm) represented by black points. This value is chosen as a threshold in agreement to studies on eruptions of similar size \citep{Andronico_et_al_2008a, Bonadonna_and_Houghton_2005, Scollo_et_al_2007} and objective limitation on measurements for historical deposits.

Two examples of sampling tests are shown in  Fig. \ref{Fig2}B and \ref{Fig2}C where the Natural Neighbor (NN) interpolation method \citep{Sibson_1981} is applied to the sampled points to reconstruct the deposit all over the domain and 20 logarithmically spaced isomass contours are drawn.

\section{RESULTS}
After an eruption, tephra deposit is often the only product available to volcanologists to estimate physical quantities in order to characterize the explosive event. The eruption “size” and style is estimated by generally inferring the mass and the particle size distribution, emitted and deposited.
In this section mass (or volume) and granulometry values, calculated using "exact integration" and classical reconstruction techniques are compared with the known values used as Scenario input.
Several well-established reconstruction techniques have been tested over 2000 sampling tests.

\subsection{Erupted and deposited mass}
 
\begin{figure}[!htb]
\centering{\includegraphics[width=1.0\textwidth]{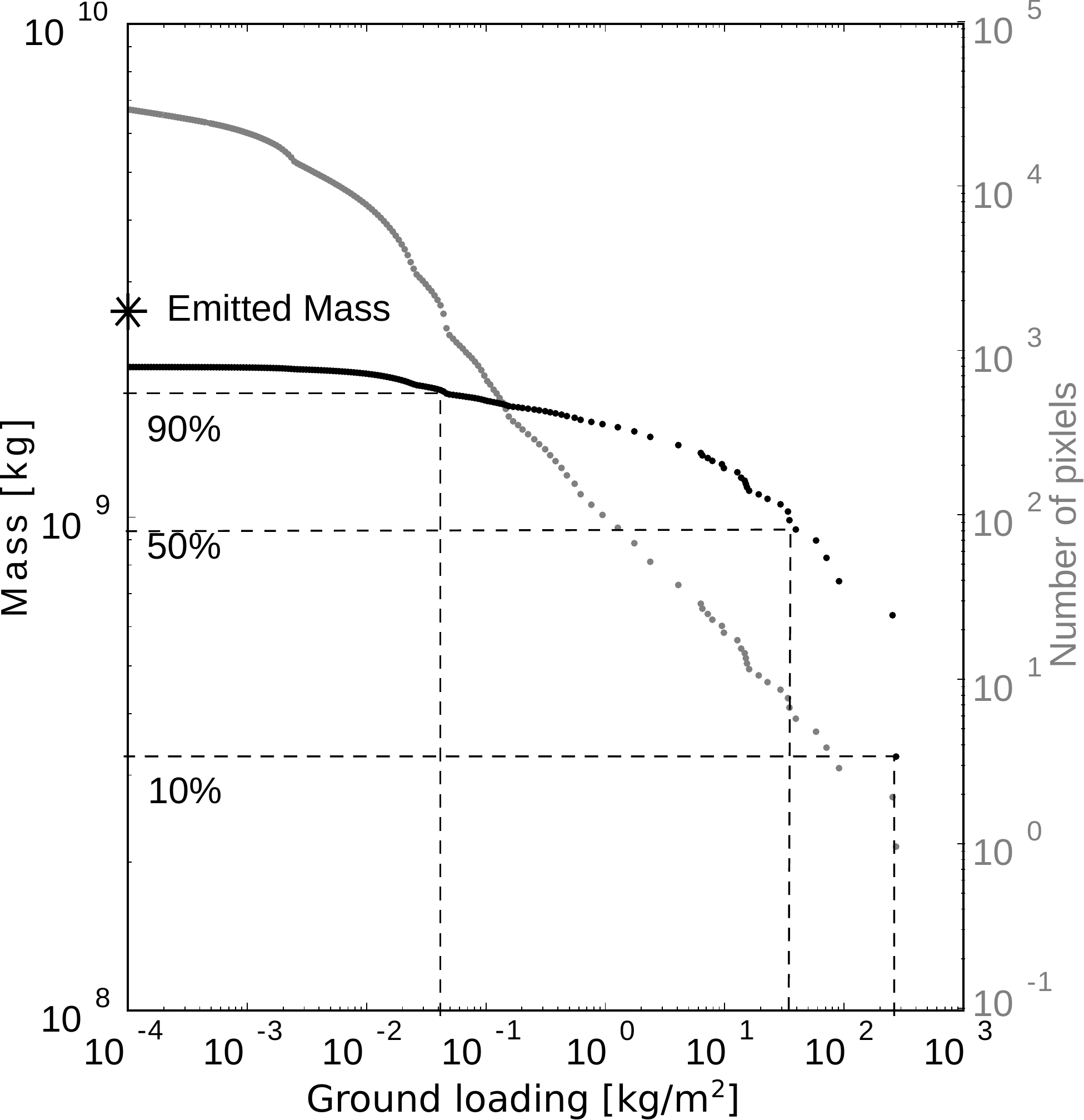}}
\caption{\label{Fig3} Cumulative mass versus mass loading. The dashed lines denotes the $ 10\%$, $ 50\%$ and $ 90\%$ of the total deposited mass and the corresponding ground loading. 
The black star on the y-axis corresponds to the emitted mass equal to $ 2.5\cdot 10^9$ kg. In gray is expressed the cumulative pixels number calculated for a given mass loading.}
\end{figure}

Here, as already stated, we first perform an “exact integration” of the ground loading produced by the numerical simulation on each cell of the computational domain. Avoiding intermediate steps as interpolation techniques and fitting functions we reduce all the subjective choices. 
In this way, the simulated deposit is analyzed to quantify the exact amount of emitted material falling in the considered domain of 250$\times$255 square cells after 24 hours. 
In Fig. \ref{Fig3} cumulative mass and cumulative number of cells vs. ground loading are plotted (dots).Total emitted mass, equal to $ 2.5\cdot 10^9$ kg, is reported on the left y-axis, and this reveals how the exact calculation of deposited mass is lower than the emitted one.

In particular the deposited mass corresponds to about $ 82\%$ of the emitted one even though the calculation has been obtained considering ground loading values down to $10^{-4}$ kg/m$^2$. 
In these figures we also plotted, with black dashed lines, the $ 10\%, 50\%$ and $ 90\%$ of the total deposited mass and the corresponding ground loading which is, respectively of $ 3.0\cdot 10^2$ , 30 and $ 4\cdot 10^{-2}$ kg/m$^2$. 

\subsection{Comparison with reconstruction techniques}

All previously introduced techniques (Pyle, Power Law and Weibull)in \ref{Erupted mass} have been applied to different sampling tests together with a NN integration as shown in two examples in Fig. \ref{Fig3}A and \ref{Fig3}B.

\begin{figure}[!htb]

\centering{\includegraphics[width=1.0\textwidth]{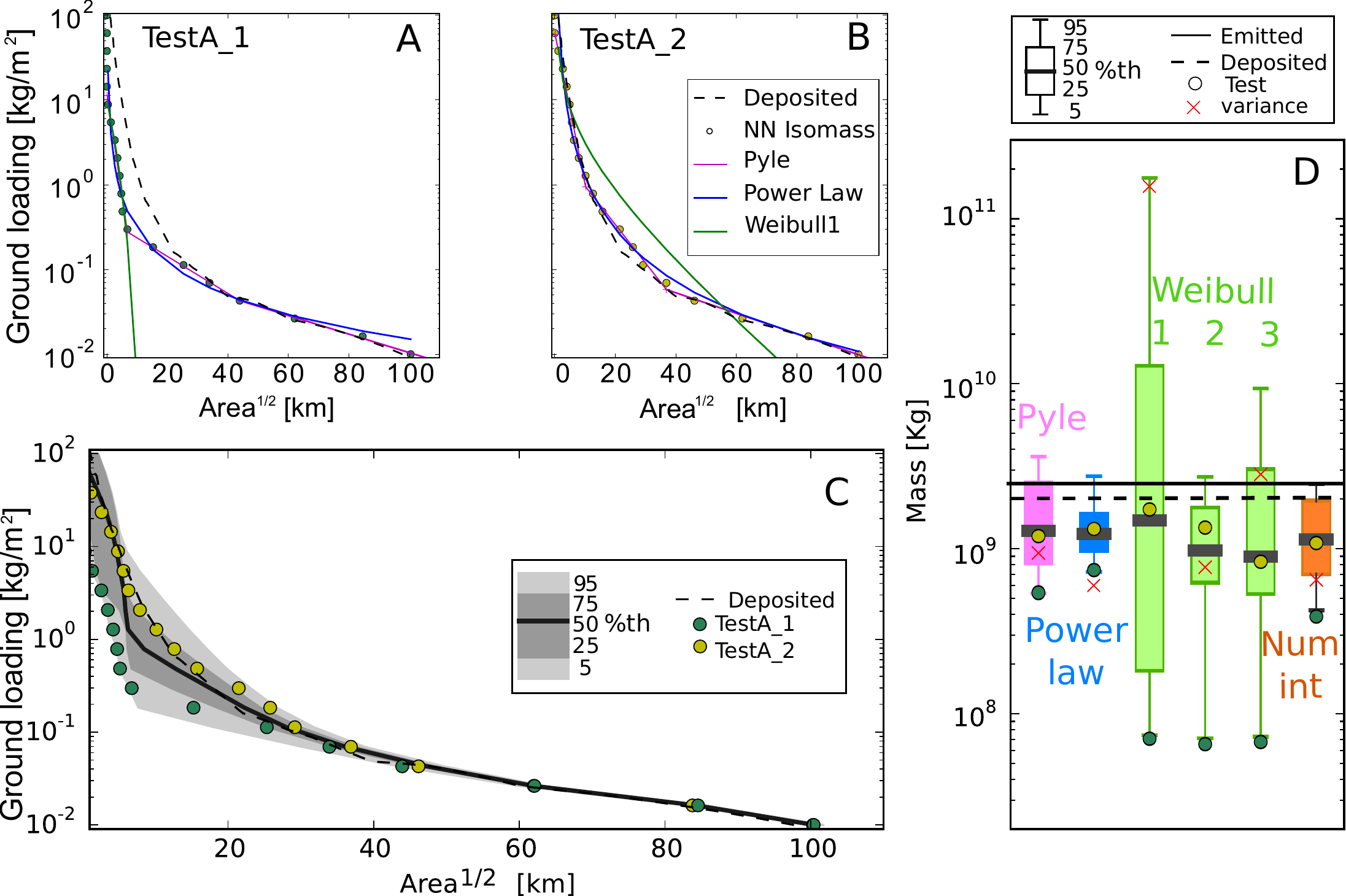}}
\caption{\label{Fig4} Semi-log plot of loading mass (kg/m$^2$) versus sqrt Area (km) for the tests presented in Fig. \ref{Fig2} (TestA\_1 in subplot A and TestA\_2 in subplot B). For each test colored circles represent the values calculated from the reconstructed deposit, whereas dashed line using the entire dataset. Pyle (magenta), Power law (blue), and Weibull (green) best fit functions are shown. 
C) Dashed line represents the deposited mass calculated by the “exact” integration over the domain. Green an yellow circles correspond to the above tests. Dark line indicates, for each ground loading, the area representing the median (50\%ile) of the isomass area values obtained for the 2000 sampling Tests. Gray and lightgray area represent respectively the $ 25^{th} - 75^{th}$ and the $ 5^{th}- 95^{th}$ percentiles intervals of calculated isomass areas. 
D) Median values, of the mass estimated for the 2000 sampling Tests obtained with Pyle, (semilog), Power law (loglog), Weibull 1 2 3 methods ($ 1/T_i, 1/T_i^2,1$ ) are reported with thick gray line. In addition to all these techniques also a numerical integration (Num Int) of the reconstructed deposit using the NN is presented (orange). The $ 95^{th} - 5^{th}$ intervals are also reported. As a reference the dashed line represents deposited mass whereas continuous black line the emitted one.}
\end{figure}
In the bottom panel of Fig. \ref{Fig4} are plotted the corresponding values of the mass loading for each of the contours versus the square root of the area enclosed by the isomass for all the 2000 sampling tests (colored dots). 
Here the results of a statistical analysis are plotted. As a reference also the two selected tests of Figure 3 are reported with green and yellow circles. Dark gray line represents the median of the isomass squared area. Whereas the gray and light gray regions  are respectively the $ 25^{th}- 75^{th}$ and the $ 5^{th}- 95^{th}$ percentile. 
 
Reconstructed isomass area values (Fig. \ref{Fig4}C) are underestimated on average for ground loadings larger than $ 10^{-1} $ kg/m$^2$ , which is where most of the deposited mass is concentrated (Fig. \ref{Fig3}). On the contrary the sampling tests, on average, overestimate the areas for ground loading values smaller than $ 10^{-1}$ kg/m$^2$.

Finally, using the different functions presented in Table \ref{Table1}, we fit ground loading values as function of squared area values and the best fitting function is integrated to calculate the deposited mass, as shown in Fig. \ref{Fig4}A-B. Black dashed line represents the mass loading vs squared area plot, obtained with the exact integration over the whole domain, as done for the plots in Fig. \ref{Fig3}. 
In  Fig. \ref{Fig4}D the estimate total mass is presented in a synthetic way, expressed for each of the techniques by the average values (averaged over the 2000 tests) with the $ 5^{th}$ and $ 95^{th}$ percentiles and the standard deviations. 
For an immediate comparison the values of the emitted mass (solid line) and exact deposited mass (dashed line) are also reported. 
All the techniques except Weibull3, on average, underestimate both the deposited and emitted mass. 
Weibull3, on average, overestimates total mass up to 1 order of magnitude, and in general using the Weibull functions a large variability in the results is produced. 
Due to this high variability, \citet{Bonadonna_and_Costa_2013} suggests to restrict the range of the Weibull parameters as function of the Volcanic Explosivity Index, obtained using Pyle and/or Power law methods. 
Preliminary tests carried out with this additional step show better results, but still presenting a quite wide variability in the reconstructed total mass values. 
Recently, different techniques have been introduced \citep{Dagitt_et_al_2014, Biass_and_Bonadonna_2011, Biass_et_al_2013} in order to quantify the interval of confidence of the fitting parameters and consequently of the estimated volume. 
In the auxiliary material is presented a study based on a Monte Carlo method showing how estimated intervals of confidences are not representative of committed errors.

\subsection{Deposited Granulometry}

In analogy with the previous section, here we first compare the emitted granulometry EM$_i$ with the deposited (DM$_i$) obtained from an exact integration of the simulated deposit over the 24 grain size classes.
Afterwards, results are compared with the TGSD obtained using the commonly adopted Voronoi tessellation method.

\begin{figure}[!htb]
\centering{\includegraphics[width=1.0\textwidth]{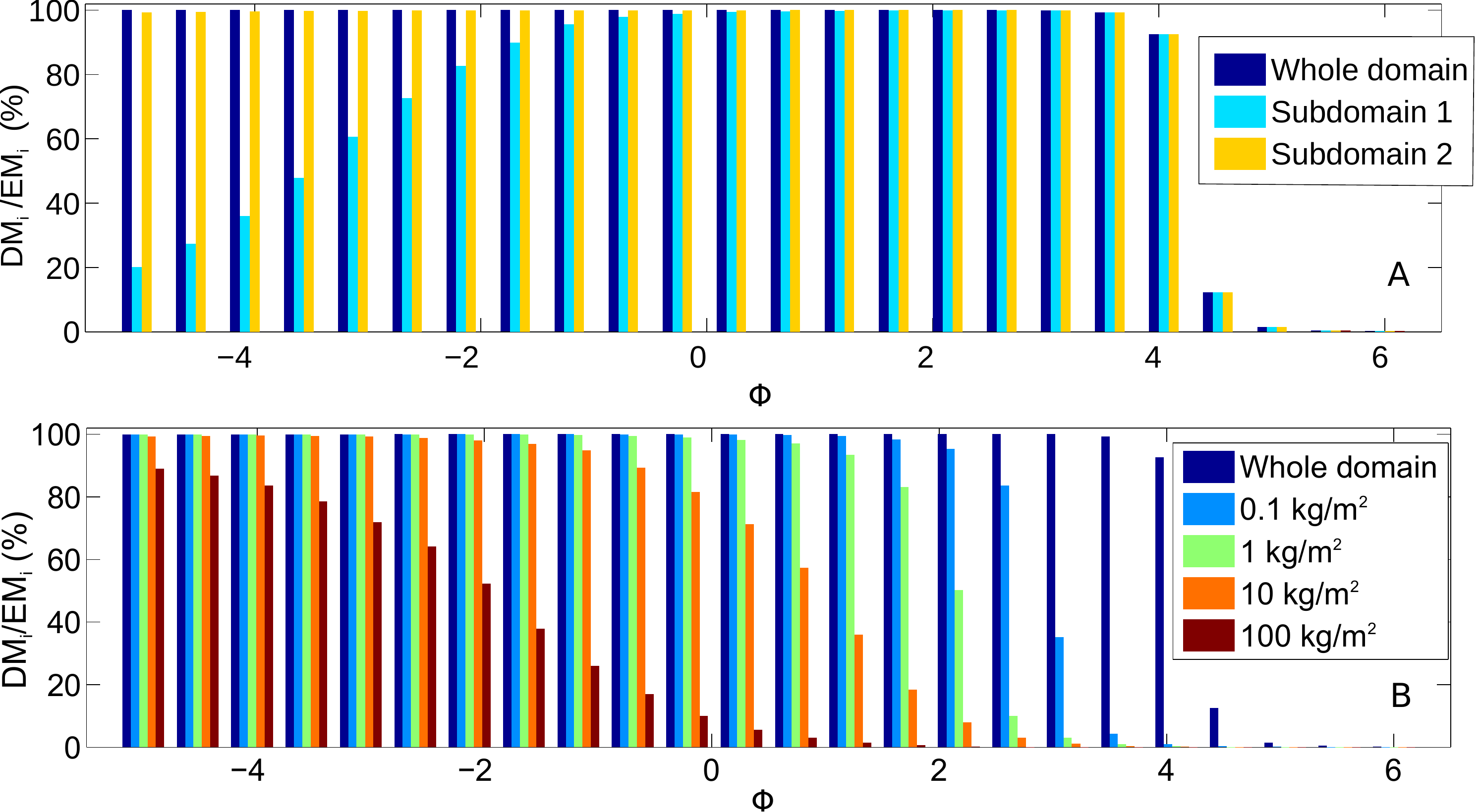}}
\caption{\label{Fig5} A) Deposited masses ( DM$_i$ ) expressed as percentages of the emitted ones ( EM$_i$ ) for each simulated grain-size. Different colors correspond to different domains, over which deposited mass has been calculated. The blue bars correspond to the whole computational domain. Subdomain\_1 and Subdomain\_2 are obtained excluding a square centered on the vent with side length of 4 km (cyan) and 2 km (yellow), respectively. B) The ratio between the deposited and emitted mass is plotted as function of minimum ground loading considered. Different colors correspond to different thresholds: 0.1 (light blue), 1 (green), 10 (orange) and 100 (brown) kg/m$^2$. The dark blue corresponds to the whole computational domain (i.e. no threshold).}
\end{figure}

Results are shown in Fig. \ref{Fig5}A, where deposited mass values obtained with the exact integration over four spatial domains (represented by different color bars) are reported in a bar plot as percentages of the emitted mass. 
Blue bars correspond to the deposit integrated over the whole computational domain. 
Other bars represent smaller domains, obtained excluding two squares centered on the vent with side length of 4 (Subdomain\_1) and 2 km (Subdomain\_2) and they can be seen as representative of samplings performed excluding very-proximal regions around the vent. 
Deposited mass (DM$_i$) (Fig. \ref{Fig5}A), down to $\phi \leq 3$, reaches 100\% of the emitted one ( EM$_i$ ). 
For $ \phi=4$ this percentage reduces to about 90\% and abruptly down to 5\% for $ \phi=5$. 
Finest simulated class is almost absent in the computed deposit. 
Results obtained excluding the material deposited inside a square centered on the vent with a side smaller than 2 km (yellow bars) do not significantly differ from those obtained considering the entire computational domain  \citep{Andronico_et_al_2014}.
On the other hand, on Subdomain\_1 (cyan bars) all the classes larger than $ \phi=0$ are strongly underestimated in the deposit, and  DM$_{tot}$ is only 68\% of  EM$_{tot}$ . 
Largest simulated class loses about $80\%$ of the initial mass. 
As expected, the deposited mass estimation strongly depends also on the minimum ground loading (hereafter referred as threshold) considered for the integration: the higher the threshold, the larger the underestimation of deposited mass. 
Fig. \ref{Fig5}B presents for each simulated grain-size the deposited mass DM$_i$ when different thresholds are considered. Classes in the $\phi$ range interval $[1,2]$ are underestimated of about 5-10\% considering a  0.1 kg/m$^2$ threshold (cyan bars); these percentages rise to 70-90\% when 10 kg/m$^2$ thresholds are considered (orange bars in Fig. \ref{Fig5}A). 
We can notice that for our scenario a large amount of the emitted mass for classes $\phi>4 $ is advected out the domain.
\begin{figure}[!htb] 
\centering{\includegraphics[width=1.0\textwidth]{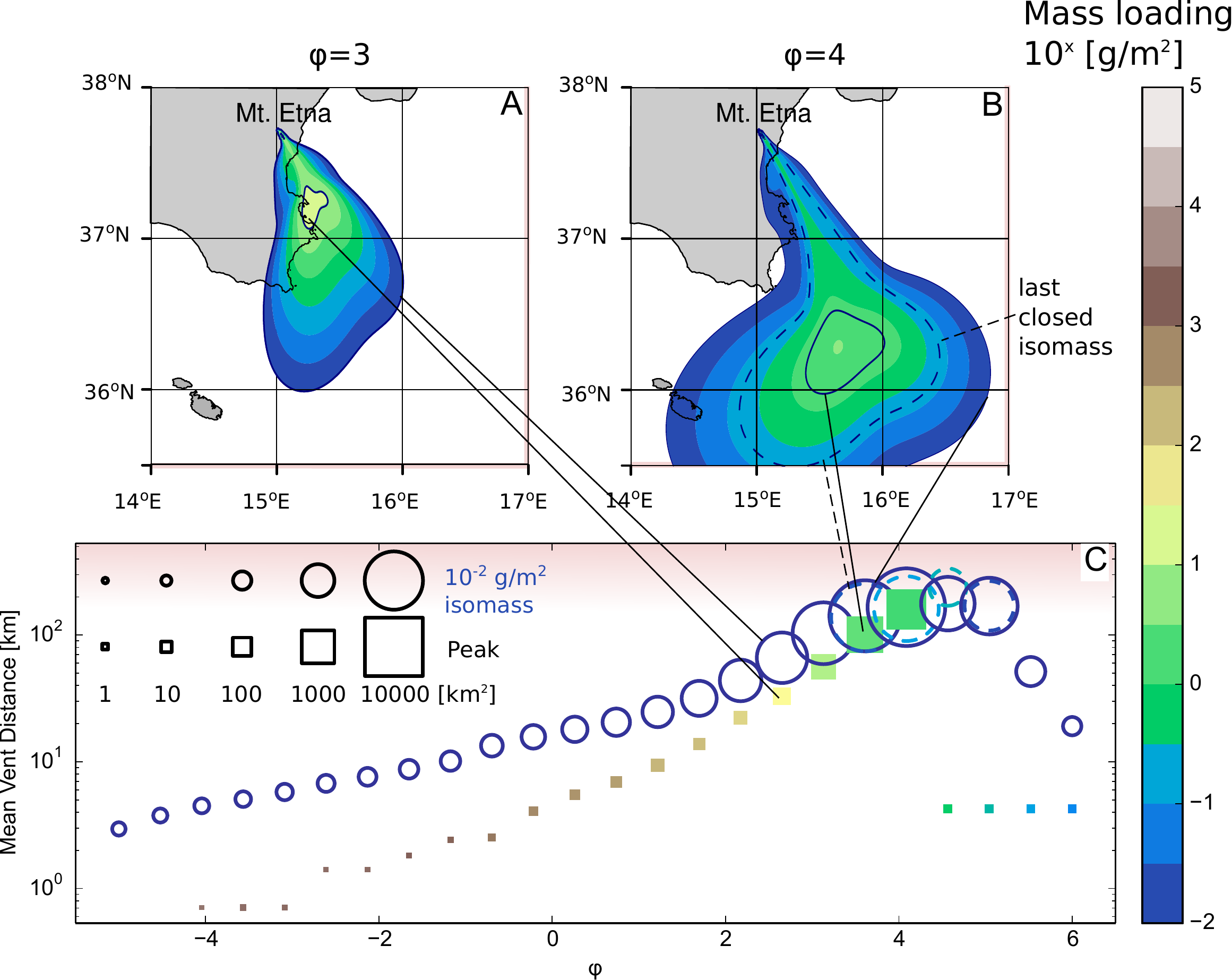}} 
\caption{\label{Fig6} Computed ground loading contours [g/m$^2$ ] at the end of the simulation for different classes: A) $\phi = 3$  and B) $\phi =4$.  Black isomass lines enclose the area with loading larger then the 50\% of the maximum (“peak”). The dashed line in B represents the last closed isomass.
C) Mean distance from the vent calculated for each class averaging the cells with ground loading larger than the 50\% of the maximum  (square), inside the 10$^{-2}$g/m$^2$ isomass (circle) and the last closed isomass (dashed circles). Color represents the cells corresponding ground load [g/m$^2$] and the dimension of the symbols the corresponding area [km$^2$]. The pink area on the y-axis of panel C denotes the distance at which ash particles reach the domain border. } 
\end{figure} 

In order to better highlight the depositional patterns for particles of different size, Fig. \ref{Fig6} shows the computed ground loading contours $[$g/m$^2$] at the end of the simulation for different classes: A) $\phi = 2$  and B) $\phi =4$.  For each class, the black isomass line encloses the area with mass loading larger than the 50\% of the maximum ground loading over the domain (max$_{cells}$(DM$_i$(cell))/2). In this way we quantify for each class both the spreading of the deposit and the amount of mass close to the peak location. This information is reported in Fig. \ref{Fig6}C, where markers size corresponds to the area enclosed by the two isomass curves. In addition, the mean distance from the vent is calculated for each class averaging the cells with ground loading larger than the 50\% maximum (square) and as a reference inside the $10^{-2}g/m^2$ isomass (circle). If this last one is open the last closed isomass is shown with a dashed circle. Colors represent the cells corresponding to ground loading $[$ g/m$^2]$ and the dimension of the symbols the corresponding area $[$ km$^2]$.  Pink area on the y-axis of panel C denotes the distance at which ash particles reach the domain border.
Smaller particles, due to the low settling velocity, are carried farther and diffuse more by the action of the wind. 
These two effects are respectively visible on the increase on the mean peak distance from the vent and on the increase of the corresponding cells number with the consequent reduction of the max ground loading for the finer classes. 

As we have already shown the effect of the domain restriction is drastically reducing the amount of deposited mass for finer classes. Here this is well visible for classes smaller than $\phi=4$ where the peak is leaving the domain border.
As consequence only the finer material falling off the margin of the column is found.  
The cutoff effect due to a limited domain is not easily overcoming, in fact the area interested by the fallout of particle finer then $\phi=4$ in the case of the described Scenario is larger than 10$^6$ km$^2$.
However the spatial distribution of small class is more homogeneous due to the larger diffusion, this is visible from the reduced distance between the last closed isomass (circles) and the peak (square).
As direct consequence, for a smaller class, the error committed using the interpolation techniques for reconstruct DM$_i$ is small. 
Since column's height depends minimally on the emitted GSD, we can conclude that for each class the peak distance from the vent is minimally depending on the emitted percentage. 
Similar results have been found using analytical model for mono-disperse plume released from height altitude \citep{Tirabassi_et_al_2009}.
Because the peak distance, for fixed column height is not depending on the emitted mass but only on the atmospheric condition, allow us to easily estimate it.
This would help to find the best region to sample, in order to better reconstruct the eruptive scenario.

\subsection{Comparison with the Voronoi technique}
We test here Voronoi technique described in section \ref{Granulometry} for reconstructing TGSD. The main requirement to apply Voronoi method is to fix the deposit extent, adding to the original data set also deposit zero values.
 We remark that fixing zero value is a strong assumption because, as shown in Fig. \ref{Fig6}, zeros are not the same for each class.
For our purposes, here we fix the zeros of the simulated deposit to correspond to the 10$^{-2}$ kg/m$^2$ isomass contours (about 0.01mm). 
As shown in Fig. \ref{Fig5}, finer particles estimation strongly depends on the thresholds defining the external boundary of the deposit, and therefore by the zero position. 

\begin{figure}[!htb]
\centering{\includegraphics[width=1.0\textwidth]{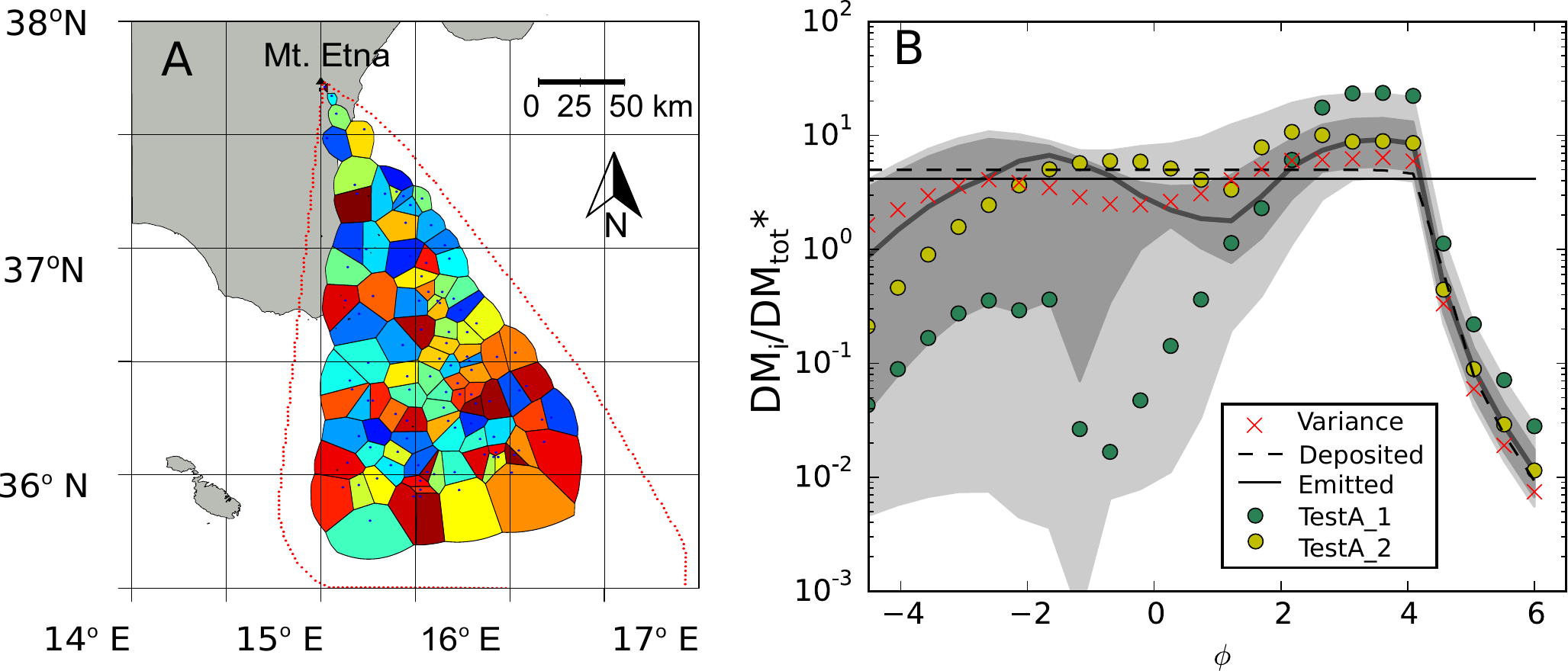}}
\caption{\label{Fig7} A) Example of a Voronoi tessellation built on sampling TestA\_1. Red crosses represent the assumed zero values, necessary for Voronoi method. Only cells with corresponding value different than zero are plotted. B) the results of the statistical analysis conducted on 2000 sampling tests are shown. For each class the median (gray line) of the mass fractions (DM$_i^*$/DM$_{tot}^*$ ) are reported. Gray area represent the interval of confidence between the $ 95^{th}- 5^{th}$ (light gray) and the $ 25^{th}- 75^{th}$ percentile (gray). Colored dots correspond to the two example tests. As a reference the mass fraction computed from the all deposit (dashed black line) and the emitted mass fraction are also reported  (black line). }
\end{figure}
Fig. \ref{Fig7}A shown as an example the Voronoi tessellation applied to one of 2000 sampling tests. In Fig. 8B the statistical results obtained applying Voronoi technique to over 2000 sampling tests are presented. This is done by plotting, for each class, the reconstructed deposited mass obtained with the Voronoi technique (DM$_i^*$), expressed as a percentage of total reconstructed mass ( DM$_{tot}^* = \sum_i$ DM$_i^*$ ) as illustrated for the two test examples presented in Fig. \ref{Fig2} (plotted with green and yellow circles). Dark gray line represents the median of the reconstructed granulometry obtained with Voronoi technique. Whereas the gray and light gray area are respectively the $ 25^{th}- 75^{th}$ and the $ 5^{th}- 95^{th}$ percentile intervals. 
These values are compared with emitted (black solid line) and deposited (dashed line) total grain size distributions. 
Standard deviations of the percentages reconstructed with Voronoi method are also reported for each class with red crosses.
For TestA\_1, the TGSD presents a multi-modal distribution, which is neither in the emitted granulometry nor in those obtained with the exact integration over the whole deposit (Fig. \ref{Fig5}), representing thus an artificial effect of the sampling, and/or the reconstruction technique.
 
One possible explanation for this is related with the fact that different classes, having different settling velocities, will generate a peak in the deposit at different distances form the vent. Indeed, if a sampling point is close to a peak of a specific class, this maximum value of the deposit will be extended over the entire Voronoi cell. This could have the final effect of overestimating the reconstructed class mass and, consequently, to cause a peak in the GSD. For particle with $\phi \geq 3$ the peaks is located beyond the domain as we can see in Fig \ref{Fig6}. Due to the diffusion smaller classes present a more uniform distribution on the ground and a smaller maximum load. For this reason also a linear method is well approximating the deposit mass.

Peak, associated with the samples and the reconstruction technique, is in the relative amount of mass and not in the absolute value, and thus could be also due to an underestimation of the other classes. 

For the classes corresponding to the peak a larger gap is present between the $5^{th}$ and the $95^{th}$ percentiles (light-gray area). A large variability in the results obtained for the different sampling tests is also indicated by the large standard deviation. 
Conversely, for finer particles, a better accordance between values obtained with exact integration and  values reconstructed with Voronoi technique is found. 
For these classes, also the gap between the $5^{th}$ percentile and the $ 95^{th}$ percentile and the standard deviation are small, and thus the reconstruction is more stable with the respect to the choice of sampling points. 
To quantify the error associated with the reconstructions obtained for each sampling test, a mean relative error averaged over the classes is defined as 
$$ err_{rel}=\frac{1}{N_{ class}}\sum_{i=1}^{N_{class}} |\frac{DM_i^*/DM^*_{tot}- DM_i/DM_{tot}}{DM_i/DM_{tot}} | .$$ 
The average of the relative errors is larger than 0.8 for both scenarios and 95\% of the sampling tests have a relative error larger than $0.35$, denoting that for the most of the test a large error is associated with the TGSD reconstruction.  

\section{DISCUSSION}
Analysis presented in the previous sections confirms that inferring quantitative data from the deposits can be very difficult and entails misleading results. Estimates values are far from representing a full picture of the initial eruptive condition at the vent. 
On one hand, the reconstruction of the total mass and the granulometry with an exact integration over the considered computational domain (an area larger than 50000 $km^2$ for the study presented here) shows that a large amount of the emitted mass is not found on the ground. 
On the other hand, the information present in the deposit is further degraded when measured by using a finite number of sampling values and then processed by reconstruction methods.
In particular,  informations on coarser particles are partially lost when  very proximal area is excluded (Fig. \ref{Fig5}A).
In addition, due to the finite area over which field samplings are performed, here simulated by a finite computational domain, and the larger area over which the dispersal phenomenon takes place, the information on the finest material is only partially reconstructed (Fig. \ref{Fig5}B). 
A combination of these two effects, here obtained considering a ground loading larger than  0.1  kg/m$^2$ (roughly corresponding to 0.1 mm) and excluding an area within about 2 km from the vent, causes an underestimation of the deposited mass up to about  50\%. 
This reflects in an almost uni-modal TGSD, reconstructed from the deposit, where the tails are both depleted with respect to the initial ones. 
We remark that the results and the underestimations discussed so far are not an effect of the procedure, since the integration is exact, but only of the choice of the integration domain, although quite extended in this study. 
In this case, the two constraints on the integration domain adopted to figure out the TGSD, i.e. ground loading > 0.1 kg/m$^2$ and distance from the vent $ > 2$ km, can reflect a reasonable minimum value sampled in the field and the objective difficulties in reaching areas very close to the vent. 
Despite the strong assumption of a uniform distribution at the vent, the deposited GSD looks similar to those reconstructed on the base of field studies for similar real eruptions at Mt. Etna (\citep{Andronico_et_al_2008a},\citep{Scollo_et_al_2007}. 
Results are obtained for a $\phi$ range spanning from -5 to 6, but the same investigations, considering finer particles, would produce similar outcomes.
Deposited mass, obtained by an exact integration, always results in an underestimation of the emitted mass EM.
This highlight the need for some correction factors to fill the gap between the estimated and the emitted eruptive parameters.
In order to generalize our results to other scenarios we conduct a sensitivity study varying the exit velocity and the radius of the column over 2400 combinations.
\begin{figure}[!htb] 
\centering{ \includegraphics[width=1.0\textwidth]{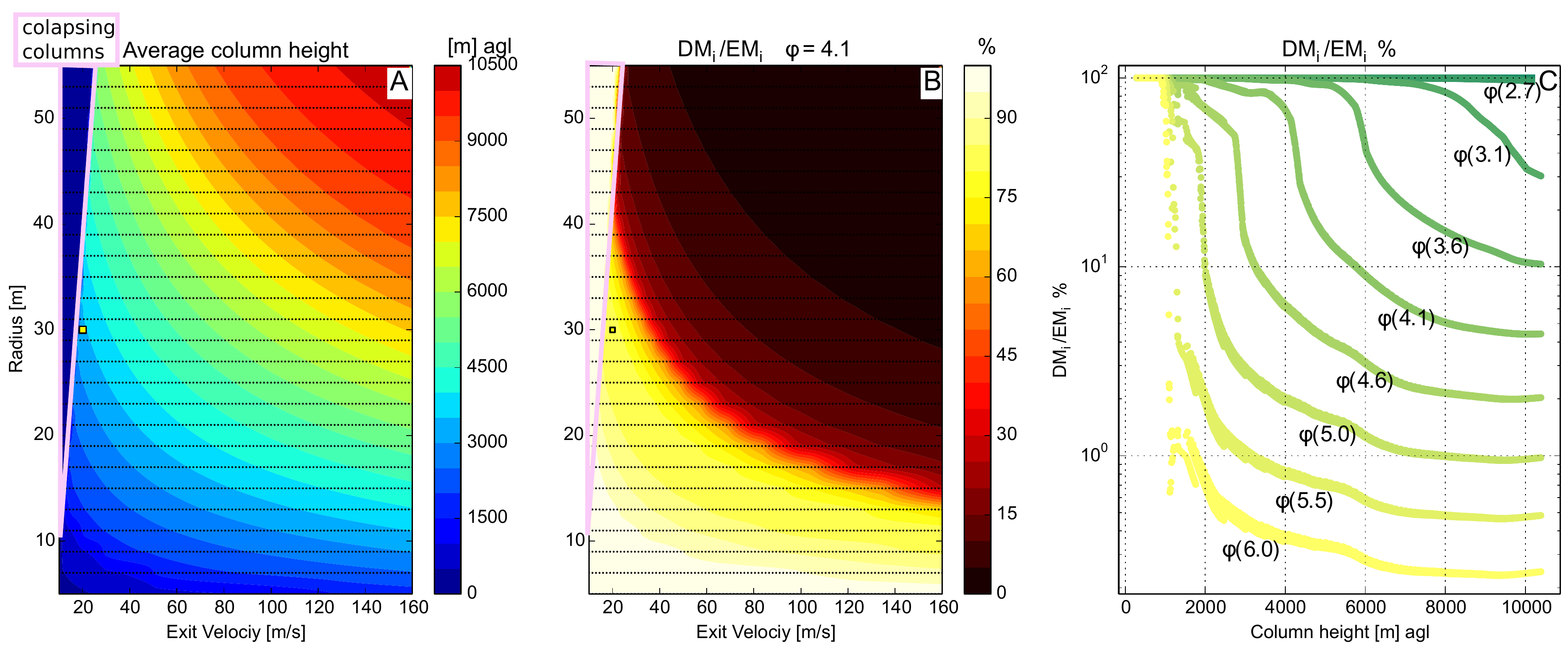}} 
\caption{\label{Fig8} A) Averaged column height over 6 hours for different scenarios.
Each scenario is the result of a run made using different initial velocities and radii for the eruptive column. The initial velocity ranges between 10 m/s and 160 m/s, and the radius between 5 m  and 55 m, resulting in a mass flow rate between  6.03 $\cdot$ 10$^{7}$ and 1.15 $\cdot$ 10$^{11}$ kg/h.  The yellow square represents the initial condition of the scenario described in the previous sections, with v=30 m/s and r=18m. The color surface is obtained interpolating the results over 2400 simulations (black dots). Similarly, in B, a color plot shows the percentage of deposited mass over the emitted (DM$_i$/EM$_i$) for $\phi$=4. In Fig \ref{Fig8}C the deposited mass (DM$_i$) for different class sizes in the $\phi$ interval [2,6] is expressed as a percentage of the emitted (EM$_i$ ).
}  \end{figure}

Fig. \ref{Fig8}A shows the average column height calculated over six hours for different scenarios. Each scenario is the result of a run made using different initial velocities and radii for the eruptive column. Initial velocity ranges between 10 m/s and 160 m/s, and the radius between 5 m  and 55 m, resulting in a mass flow rate between  6.03 $\cdot$ 10$^{7}$ and 1.15 $\cdot$ 10$^{11}$ kg/h.  Yellow square represents the initial condition of the scenario described in the previous section, with v=30 m/s and r=18m. Color plot is obtained interpolating the results over 2400 simulations (black dots). Similarly, in Fig. \ref{Fig8}B, a color plot shows the percentage of deposited mass over the emitted (DM$_i$/EM$_i$) for $\phi=4$. If we look at the reference scenario discussed in the paper (yellow square) on Fig \ref{Fig8}B, we see that the deposited mass over the considered domain for $ \phi=4$ corresponds only to the $ 90\% \pm 5\%$ of the emitted one (see also Fig.\ref{Fig5}). In Fig \ref{Fig8}C the deposited mass (DM$_i$) for different class sizes in the interval $[2,6]$ in the $\phi$ scale is expressed as a percentage of the emitted (EM$_i$ ). This plot allows defining a correction factors for the other classes, and more general for different eruptive scenarios. In this way we can relate the deposited TGSD with the emitted one. 
 Before applying the method we need to be careful considering the uncertainty related to the reconstruction techniques applied to samplings dataset. 
This is highlighted in Figs. \ref{Fig4}D and \ref{Fig7}B where the erupted mass (EM$_{tot}$) and the GSD are obtained respectively with Pyle, Power law and Weibull methods and with the Voronoi technique. 
In particular, despite the large number of samples considered for each sampling test (100 points could be considered a very well sampled deposit), the erupted mass values obtained show a large variability and, for a significant percentage of tests, an overestimation (Fig. \ref{Fig4}D). 
In general, the Pyle and Power law fitting methods seems to produce more stable results than the Weibull function, providing a smaller standard deviation within the 2000 sampling tests performed, and making more reliable the application of a correction factor. 
Also when Voronoi technique is used, TGSD presents fictitious peaks in the resultant distribution (Fig. \ref{Fig7}B) and a large variability in the results are produced depending on the choice of the sampling points with regard to locations.
As already discussed, the grain size distribution is expressed as fractions of the total mass and not as absolute values of deposited mass for each class. 
In this way the error of a single class affects all the others as individual mass fractions depend on total mass.
In this work measurements are considered not affected by errors.
Ultimately, measurements errors could have the effect of a further deterioration of the information present in the deposit, decreasing the margin of confidence in the results. 

Depositional models tend to underestimate ground loading in the proximal region and consequently overestimate distal deposited mass. This is mostly due to processes neglected by the models, for example aggregation \citep{Taddeucci_et_al_2011} and gravity currents derived by instability \citep{Manzella_et_al_2015}.
In both cases finer particles are settling down with a higher settling velocity and consequently the amount of finer particles deposited is larger.
On the other side smaller particles aggregated on a larger one modify the drag coefficients of the aggregate, reducing its settling velocity. As a consequence larger particles involved in the aggregation can travel longer.

\section{CONCLUSION}
In order to better understand a deposit footprint and its thoroughness in quantifying eruptive source parameters like mass and GSD, a synthetic tephra deposit has been generated by using a dispersal code, and then analyzed. 
We performed this analysis by estimating the emitted and deposited values of mass and GSD by using an exact integration over the computational domain as well as using standard reconstruction techniques. 
These values have been then compared with the emitted mass fixed a priori as input for the numerical simulations. 
Exact integration showed an underestimation of the initial quantities mainly due to the finite extension of the domain, but still consistent with the physical processes occurring during the transport and the sedimentation, for example the longer distance and time-travel associated with fine particles. 

Quantitative reconstruction of eruptive conditions based on deposit sampling solely can be affected by large uncertainty, and can address result in misleading interpretation. 
Even by using a large and well-distributed dataset of sampled points (100 points) over the modelled domain the results can be affected by large errors. 

In this case, estimated mass was not representative of emitted one. Reconstruction techniques, performed over 2000 different sampling tests, showed a gap in the mass values up to an order of magnitude between the $ 5^{th}$ percentile and the $ 95^{th}$ . 
Large standard deviation values and large relative errors affected reconstructed grain size distribution obtained with Voronoi. The technique has been applied to 2000 sampling tests of 100 points, revealing the strong sensitivity of the results on the considered samples and often producing fictitious peaks in the distribution. 
In this case, an exact integration performed considering a threshold on the ground loading and excluding a proximal area (reflecting a reasonable minimum value sampled in the field and the objective difficulties in reaching areas very close to the vent), produced an almost uni-modal GSD where the tails are both depleted. 

In particular the emitted fraction of finer particles was underestimated, with consequently a bias on the associate  assessment of risk, especially for airborne.

We point out that obtained outcomes are not the image of a single eruption. Results have been extended using a sensitivity study on eruptive parameters, generalizing the analysis conducted so far depending on the eruption scale. 
Finally, this work showed how, in support of field-based studies, numerical models can represent a useful tool to better understand the complexity of the deposition process and, consequently, to manage the associated uncertainty. 

On one hand, exact integration performed over the simulated deposit, which resulted in an underestimation of emitted mass has  been used to define some correction factors for estimation of source eruptive parameters. 
On the other hand, results obtained with the exact integration provided reference values to compare the different integration techniques and assess their reliability.

In order to extend and generalize presented results other tests have been performed and presented on the auxiliary material 

\section{Recommendations}
In the following part we summarized article result in a best practice list.
In order to limit arbitrary choices on estimation process, our recommendations are:
\begin{itemize}
\item Collecting values along the major dispersion axis with a log spaced distance from the vent.
\item Collecting values along minors dispersion axis.
\item Collecting values upwind. 
\item Collecting zeros values to constrain deposit limits. 
\item Using an interpolation technique as Natural Neighbor for each class (no need of assigning zeros values as Voronoi). 
\item Integrating class by class.
\item Using correction factors to account for domain restrictions.

\end{itemize}

\section{ACKNOWLEDGMENTS}
This work presents results achieved in the PhD work of the first author (A.S.), carried out at Scuola Normale Superiore and Istituto Nazionale di Geofsica e Vulcanologia.  
The activity has been partially funded by the Italian Presidenza del Consiglio dei Ministri \- Dipartimento della Protezione Civile (Project V1). Further we should mention ARPA\_SIM-CINECA for the LAMI code runs. The manuscript benefited from fruitful discussions with Dr. A. Neri and Dr. S. Engwell.
\printbibliography[heading=subbibliography]
\end{refsection}
\pagebreak

\appendix

\titleformat*{\section}{\LARGE\bfseries\sffamily}
\section{Auxiliary Materials\\Reconstructing eruptive source parameters from tephra deposit: a numerical approach for medium-sized explosive eruptions}
\setcounter{table}{0}
\renewcommand{\thetable}{A\arabic{table}}

\setcounter{figure}{0}
\renewcommand{\thefigure}{A\arabic{figure}}
\renewcommand\theHfigure{back.\arabic{figure}}

This auxiliary material contains:  
\begin{itemize} 
 \item A sensitivity study of eruptive initial parameters. 
\item An analysis of the uncertainty for the volume estimation techniques based on a Montecarlo method.
\item A sensitivity study of a simple aggregation model 
\end{itemize} 
\subsection{Sensitivity analysis} 
In order to generalize the results presented in the article, a sensitivity study of VOL-CALPUFF has been performed. 
Tab \ref{Tab1} resumes main volcanological input parameters used to initialize the different runs. In the following we will refer to the main example presented on the article with Scenario\_A.

\begin{table}
\begin{center}\small 
\begin{tabular}{c c c c c c c} 
 \hline        & & Model input data & & &\\ 
 \hline
 
  Run    & Grain size & Particles & Particles & Exit     & Column  \\   
         & distribution &  Factor shape            &  density       & Velocity & Radius\\ 

         & $\phi$ = [-5,6]   & $\psi$ & (kg/m$^3$) &(m/s)          & (m)     \\   
\hline           
         Scenario\_A & uniform & 0.5 & 2000 & 30   & 18       \\
         
         Fig \ref{Fig4_aux} & lognormal $\sigma \in$[0,6]  $\mu \in$[-3,5]   &0.5 & 2000& 30 & 18      \\ 
         Fig \ref{psiANDrho} & uniform &[0.5,1] & [1500,2000]     & 30    & 18       \\
         & uniform & 0.5 & 2000     & [10,160]    & [5,55]       \\

\end{tabular}  
 \end{center} 
\caption{\label{Tab1}
Main volcanological input used for the sensitivity study. Square brackets represent the interval where the parameters are varied.} 
\end{table}

\begin{figure}[!htb] 
\centering{ \includegraphics[width=1.0\textwidth]{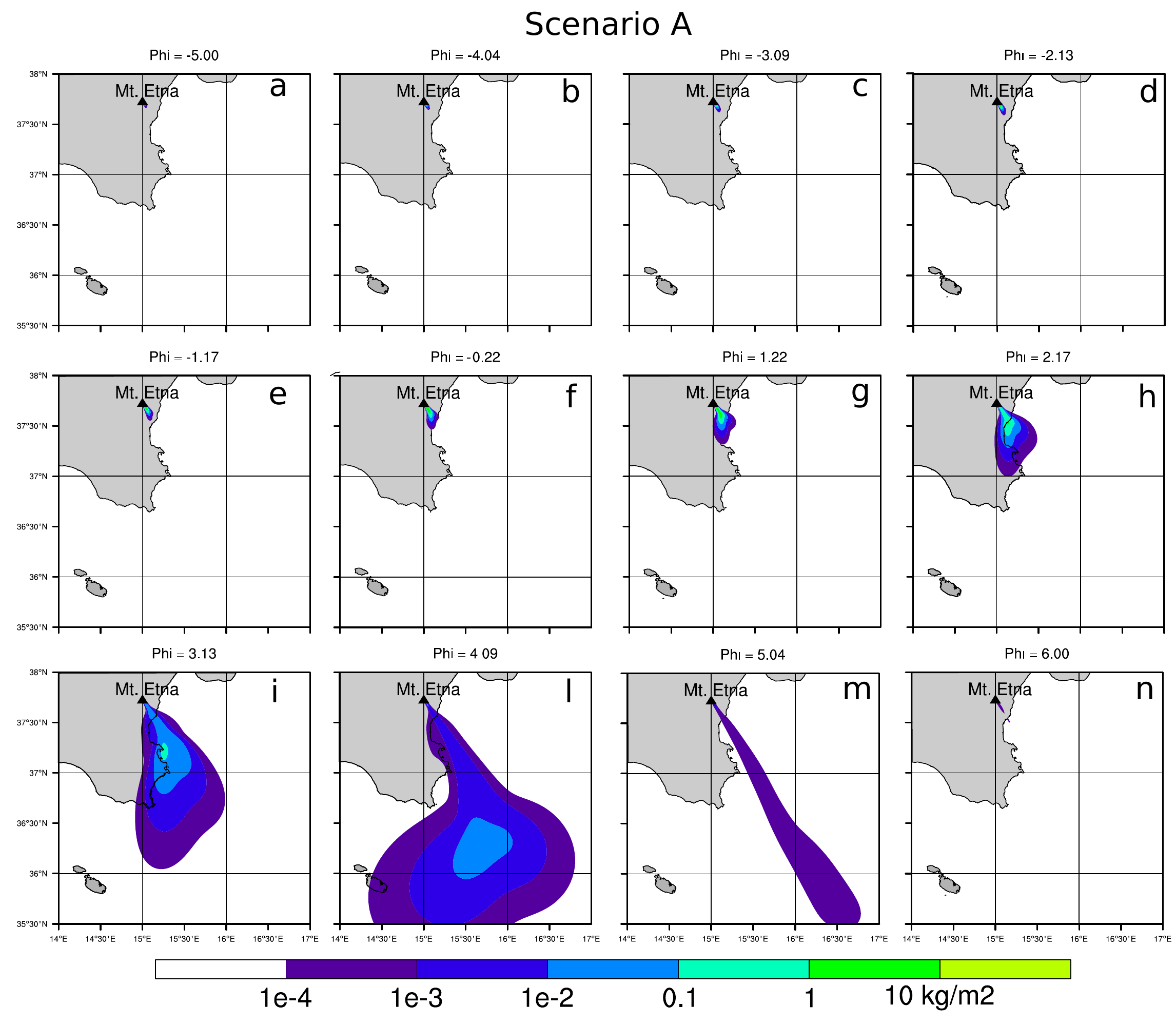} } 
\caption{\label{Fig1_aux} Computed ground loading contours [kg/m$^2$ ] for Scenario\_A at the end of the simulation for different classes: A) $\phi =-5$  ... E) $\phi = 6$.  
Isomass contours are log-spaced from 10$^{-5}$ to $10$ [kg/m$^2$ ]. In all plots the black triangle indicates the position of the eruptive vent.} 
\end{figure} 

Fig. \ref{Fig1_aux} shows a detailed deposit map for Scenario\_A, where deposit (DM$_i$) is plotted for different particle classes: A) $\phi =-5$  ... E) $\phi = 6$.
Isomass contours are log-spaced from 10$^{-5}$ to $10$ [kg/m$^2$ ].
Although smaller particles have a larger diffusion coefficient they are carried farther by the wind due to the low settling velocity. Therefore most of the finer particles ($\phi$ >4) are leaving the domain and only a few percent of the emitted mass (EM$_i$) is deposited on the ground over the considered domain. Finest simulated class ($\phi = 6$), as shown in Fig \ref{Fig1_aux}n, is almost absent in the computed deposit. Moreover, we can notice that different classes present different main axis directions on the deposit.
This segregation process is mostly due to a shear on the wind direction with the altitude. Smaller particles are more influenced by higher wind directions and consequently also their deposit trace.
This suggests to be careful when inferring the main dispersal axis and consequently when choosing sampling locations.
Wind measurements and reliable weather data are fundamental if we want to proper characterize the footprint left on the ground by an eruption.
 
\begin{figure}[!htb] 
\centering{ \includegraphics[width=1.0\textwidth]{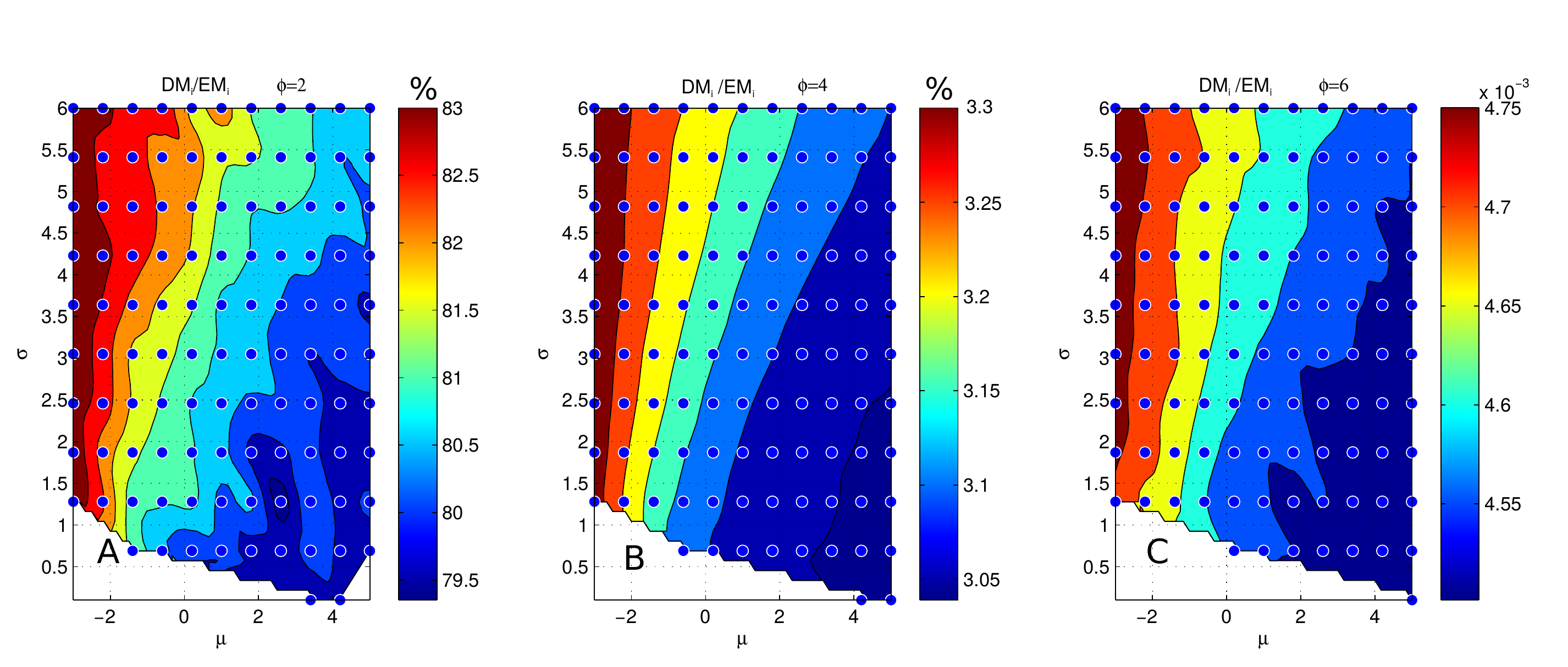} } 
\caption{\label{Fig4_aux} Deposited mass (DM$_i$) for different class sizes $\phi =2$ (A) $\phi =4$ (B) $\phi =6$ (C) expressed as a percentage of the (EM$_i$ ). A blue dot represents a run made using as emitted GSD a log normal distribution with assigned parameters $\mu$ and $\sigma$. Colored surface is obtained interpolating the results over 110 simulations. 
} 
\end{figure}

In order to generalize the results presented in the main paper, we investigate here results dependency on the emitted grain size distribution (GSD). 
Fig. \ref{Fig4_aux} shows the results of 110 runs where emitted GSD is chosen using 24 classes with a log-normal distribution described by different values of $\sigma$ and $\mu$. Each dot represents a different run and the color map is obtained interpolating the values of DM$_i$/EM$_i$ from the dots location. Results are shown for three different classes $\phi=2$ (A), $\phi =4$ (B) and $\phi =6$ (C). For each of these classes the variability of DM$_i$/EM$_i$ is really small ($3-4\%$ respect to the mean value), and for $\phi>4$ this ratio is even smaller. 
Therefore the percentage of deposited mass with respect to the emitted one is almost independent from the emitted GSD, allowing to generalize the results obtained for Scenario\_A also to GSDs different from the original one presented in the manuscript.

In addition to the analysis on the effect of the initial particles distribution, here we want to investigate the role of particle shape factor $\psi$ and density $\rho$ on the transport and deposition process. Several experimental studies have highlighted how an irregular shape of falling particles could affect their velocity toward the ground \citep{walker71, wilson72, Wilson_and_Huang_1979, riley, dellino}. As introduced in \citet{riley} the sphericity parameter $\psi$ is defined as the ratio between the projected area $A_p$ and the square of the projected perimeter $P_p$:  

$$
\psi=\frac{4\pi A_p}{P_p^2}.
$$ 
 
\begin{figure}[!htb] 
\centering{ \includegraphics[width=1.0\textwidth]{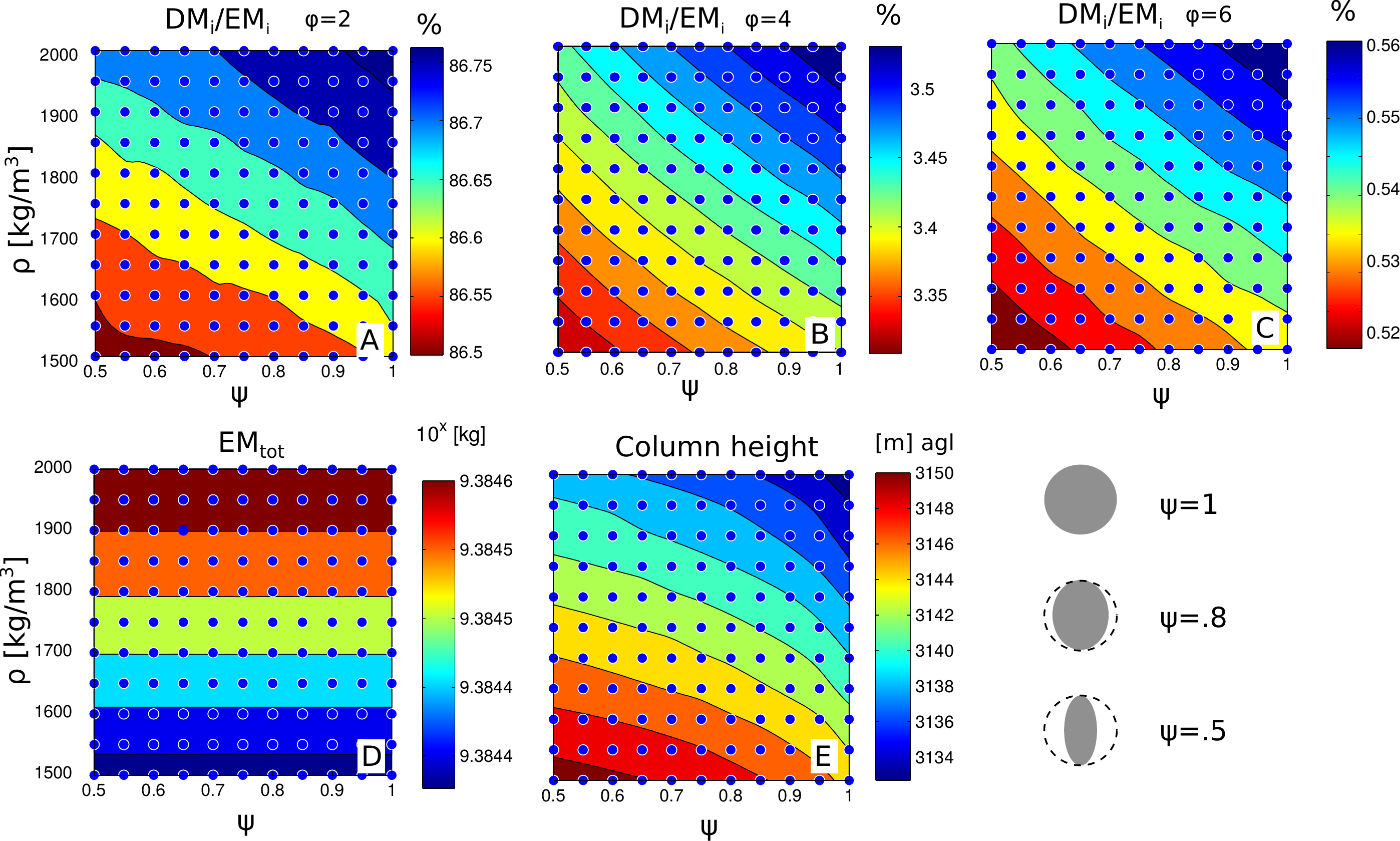} } 
\caption{\label{psiANDrho} Deposited mass (DM$_i$) for different class sizes $\phi =2$ (A) $\phi =4$ (B) $\phi =6$ (C) expressed as a percentage of the emitted (EM$_i$ ). A blue dot represents a run made using as emitted GSD a uniform distribution where particles have different factor shape $\psi \in [0.5, 1]$ and density $\rho \in $ [1500, 2000] kg/m$^3$. Colored surface is obtained interpolating the results over 121 simulations. 
D) Emitted mass and E) averaged column height. F) Schematic example of different particles with different sphericity $\psi$.}
\end{figure}

In Fig \ref{psiANDrho} deposited mass DM$_i$ for different class sizes ($\phi =2$ (A), $\phi =4$ (B), $\phi =6$ (C)) is expressed as a percentage of the emitted mass EM$_i$. Each blue dot represents a run made using as emitted GSD a uniform distribution but with different factor shape $\psi \in [0.5, 1]$ and density $\rho \in $ [1500, 2000] kg/m$^3$. In Fig \ref{psiANDrho}F is presented a schematic example of different particles with different sphericity values $\psi$. For spherical particles $\psi=1$. Colored contour plot is obtained interpolating the results over 121 simulations. In Fig. \ref{psiANDrho}D and \ref{psiANDrho}E) are plotted respectively as input response surfaces the emitted mass $EM_{tot}$ and averaged column height.  
We can notice that for a fixed $\phi$ class the ratio DM$_i$/EM$_i$ increases with increasing density and sphericity. 
In fact more irregular and lighter particles are transported farther. Besides, relative changes in the DM$_i$/EM$_i$ increases with the decreasing size: 0.3\% for $\phi=2$, 7\% $\phi=4$ and 8\% for $\phi=6$. As expected EM$_i$ (Fig \ref{psiANDrho}D) increases with particles density and is not influenced by the sphericity,
whereas denser particles produce lower column heights (see Fig \ref{psiANDrho}E). Finally, less spherical are the particles, less rapid is their vertical displacement allowing a longer staying in the atmosphere.

\subsection{Uncertainty analysis} 

In order to quantify the interval of confidence of the fitting parameters and consequently of the estimated volume, different techniques have been recently introduced \citep{Dagitt_et_al_2014, Biass_and_Bonadonna_2011, Biass_et_al_2013}.
Model misspecification on fitting process may cause large bias thus leading to incorrect inference. 
In this section we want to understand how the uncertainty in eruption volume estimation is depending on the fitting model (for example the difference between applying the Weibull model vs the exponential model) and on fitting errors. 
First of all, we want to compare the intervals of confidence for the volume estimation obtained applying the different techniques presented in the manuscript.
On \citet{Dagitt_et_al_2014} the estimation is based on a least-squares method. Relative squared error is used to measures how well a fitting function $f$ fits the data $y$. Errors are assumed normally distributed. 
\begin{figure}[!htb]
\centering{\includegraphics[width=1.0\textwidth]{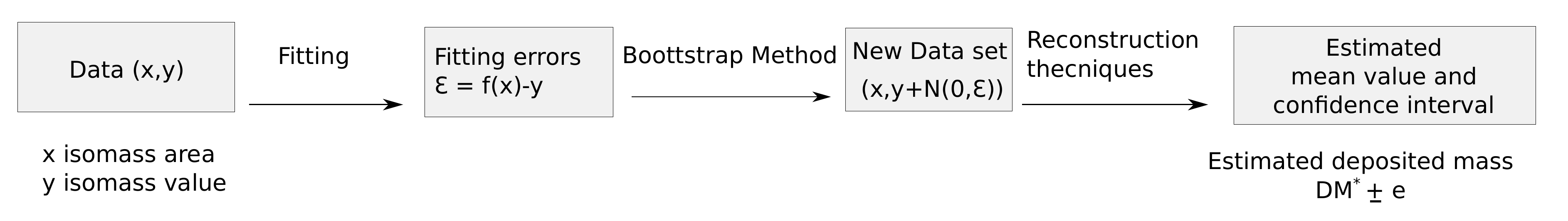}}
\caption{\label{sketch} 
Schematic representation of the bootstrap method.
}
\end{figure}

Here in a similar way of \citep{Biass_et_al_2013} we used a bootstrap Monte-Carlo method \citep{Efron_and_Tibshirani_1993} as presented in Fig \ref{sketch}. 
In order to create a large dataset, the first step is to randomly perturb the calculated isomass area.  
If errors in the isomass squared area calculation ($y$ points) are available, then the random variation in each point is determined from its given error. Otherwise, the random variation is determined from average standard deviation of $y$ points. 
A large number of random data sets are then produced and fitted, and the variance of fit results is used as the error for the fit parameters.
Even if estimations of observational errors have been produced for mass loading measurement \citep{Bernard2013}, it not trivial to predetermine the error distribution on corresponding isomass area, especially for sparse  sampling points dataset. 
\begin{table} 
\begin{tabular}{lllllll} 
 
Test  &  & Pyle & Power Law & Weibull1 & Weibull2& Weibull3\\ 
 
\hline 
 
    & Residual norm  &  0.12 &  0.41 &  1.08 &  0.5 &  0.5\\ 
 
1 A & Mean Volume     &  0.55 & 0.81 & 34.8 & 6.42 & 6.42\\ 
 
 
\hline 
 &Residual norm  &  0.14&  0.32 &  21 &  0.26 &  1.65\\ 
 2 A & Mean Volume    &  1.3 & 1.36 & 66.3 & 1.49 & 1.49\\ 
 \hline 
\end{tabular} 
\caption{\label{Tab2}Residual norm, Volume mean and Volume error estimate for all the volume reconstruction methods. Error estimation is calculated using a boot strap method. Values are expressed as a fraction of the emitted volume.} 
\end{table} 

Bootstrap method has been applied to the volume reconstruction techniques from two selected tests (TestA\_1, TestA\_2) using 1000 "perturbed" $y$-data sets. 
Analysis results are presented in table \ref{Tab2}.  
We can notice the following main facts: 
\begin{itemize} 
\item a small residual norm (i.e. a good fitting) does not guarantee a small interval of confidence.   
\item estimate error and consequently the estimate confidence interval are not representative of the real error committed.
\end{itemize} 
Moreover, comparing the three different Weibull methods, we observe that results strongly depend on the choice of the residual function, and generally the volume mean is smaller than the volume errors. 
Unfortunately, it is difficult to find a non-subjective criterion for choosing the appropriate residual function.   
Even if we can find a satisfactory fitting of the mass loading vs squared area, this does not ensure a good volume estimation when errors are also associated with area values.   

However the estimation is based on the assumption that the fitting model is valid, and does not account for error in the interpolation of the isomass contours from sparse data. 
Finally we have to stress that, to a lesser extent, the mass estimation methods proposed are also sensitive to the values and the number of isomasses. 

\subsection{Aggregation} 

As stated in several papers \citep{Carey_and_Sigurdsson_1982, Bonadonna_et_al_2011a, Taddeucci_et_al_2011, Brown_et_al_2012}, particle aggregation can significantly affect the dispersal process and the tephra depositional pattern.  
For this reason we extended our results considering a simple aggregation process. As proposed by \citet{Textor_et_al_2006} aggregation is supposed occurring within the margin of the column.  
To this aim, we adopt a model based on \citep{Cornell_et_al_1983, Sulpizio_et_al_2012}, where $10-50\%$ of emitted mass for $ \phi>2$ particles aggregates as $ \phi=2 $ ($ 250$ $ \mu m$ ) particles. As a first approximation particle density is not modified and is equals to 2000 kg/m$^3$.  
As stated by \citet{Brown_et_al_2012}), aggregates, due to the collision on the ground and during measurements  are supposed to be found as completely break apart on the deposit. 
\begin{figure}[!htb] 
\centering{ \includegraphics[width=0.6\textwidth]{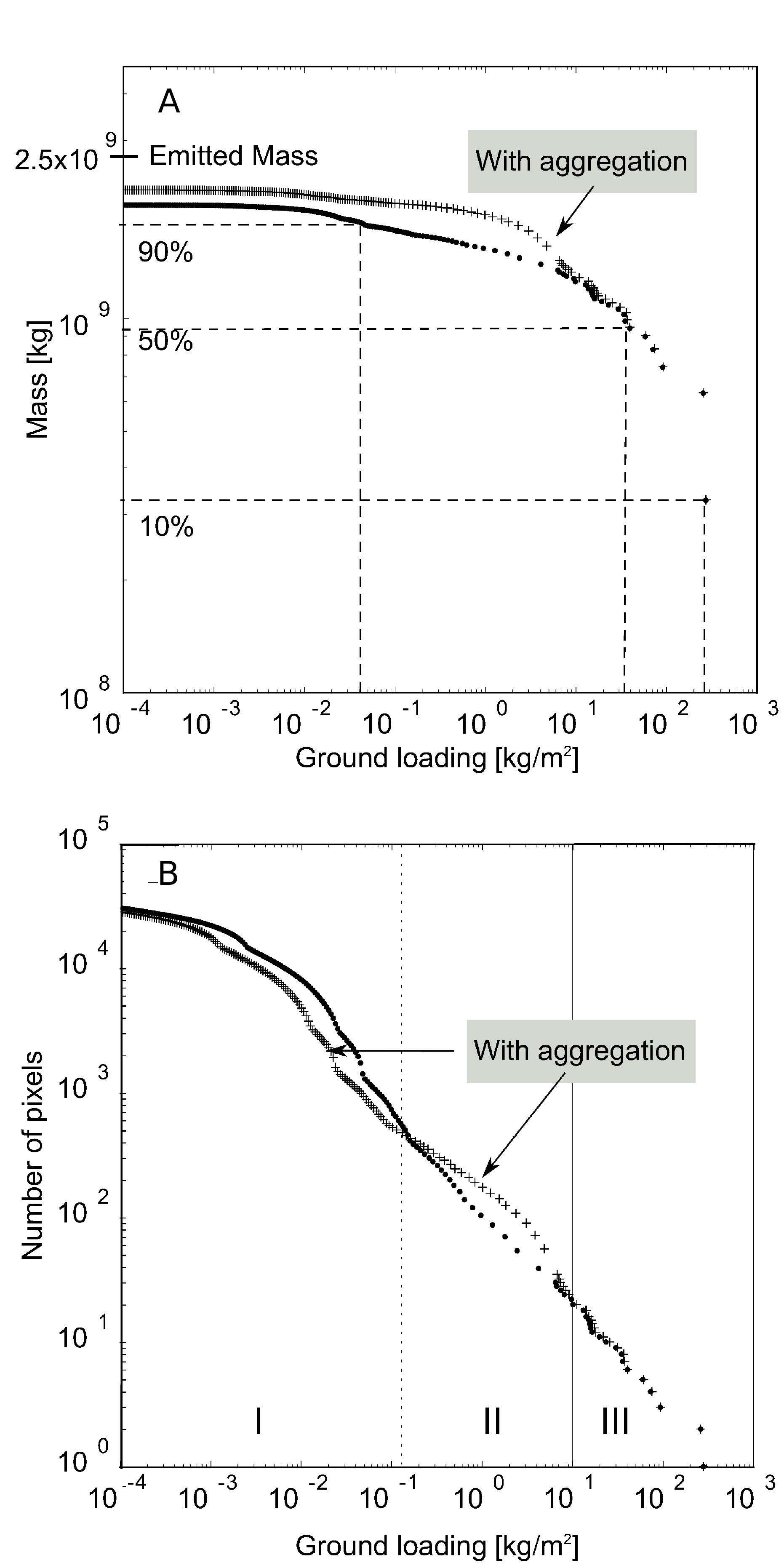} } 
\caption{\label{Fig5_aux} A) Cumulative mass versus mass loading.  
Dashed lines denote the $ 10\%$ , $ 50\%$ and $ 90\%$ of the total deposited mass and the corresponding ground loading.  
Black segment corresponds to the EM$_{tot}$ equal to $ 2.5\cdot 10^9$ kg.  
B) 
Figures shows results without (dots) or with $ 50\%$ degree of aggregation (crosses). 
We can identify three intervals in ground loading values where the crosses (simulations with aggregation) are respectively below (interval I) (at distance), above (interval II) (intermediate) or coincident (interval III) (proximal). 
} 
\end{figure}

In Fig.\ref{Fig5_aux}A  cumulative mass is plotted as a function of ground loading. Total emitted mass (EM$_{tot}$), equal to 2.5 $ \cdot$ 10$^9$ kg, is reported on the left y-axis.
This reveals how DM$_{tot}$ corresponds to about 82$\%$ of EM$_{tot}$, even though the calculation has been obtained considering ground loading values down to 10$^{-4}$ kg/m$^{2}$ over an extended domain. In these figures, for the simulations without aggregation, we also plotted, with black dashed lines, the 10$\%$, 50$\%$ and 90$\%$ of the total deposited mass ( DM$_{tot}$) and the corresponding ground loading which is, respectively of 3.0$ \cdot$10$^2$, 30 and 4$\cdot$10$^{-2}$ kg/m$^{2}$. 
Results are shown considering no aggregation (dots) or with $ 50\%$ degree of aggregation (crosses). 
When $50\%$ of aggregation is introduced in the model, a larger amount of mass is deposited within the domain, resulting in an increase of $ 10\%$ for Scenario$_A$. 
Fig. \ref{Fig5_aux}B shows cumulative pixels number calculated as a function of mass loading for simulation with and without aggregation. Three intervals can be identify in ground loading values, where the crosses (simulations with aggregation) are respectively below (interval I) (at distance), above (interval II) (intermediate) or coincident (interval III) (proximal) with the dots (simulations without aggregation).  
These intervals depend on the way aggregation is modelled (in this study on the percentages assumed to aggregate) and, above all, on the dimension of the aggregates and their fall velocities.
Most of the particles involved in the aggregation process do not fall in proximal areas.  Consequently, close to the vent, where the cumulative ground loading is larger than 10 kg/m$^2$ aggregation has a negligible effect (interval III).  Below these values aggregation reveals its main effects. In particular, in the range $[10^{-1},10]$ kg/m$^2$ (interval II) a large portion of particles falls over a smaller area with higher loadings.  
At the same time, the area corresponding to low loading values, here smaller than $ 10^{-1}$ kg/m$^2$ (interval I), is reduced.

\begin{figure}[!htb] 
\centering{\includegraphics[width=1.0\textwidth]{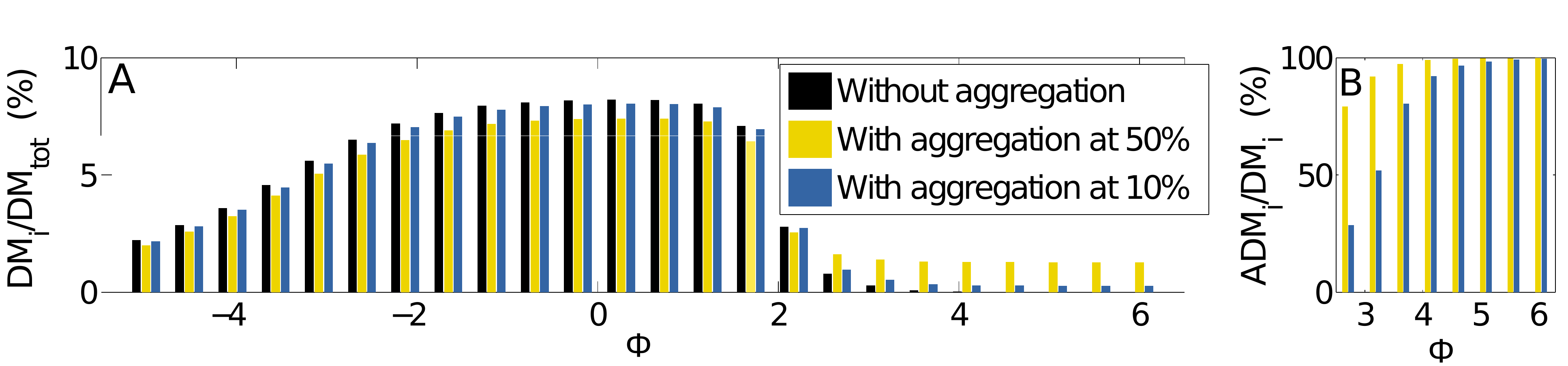}} 
\caption{\label{Fig6_aux}A) i-th deposited mass ( DM$_i$ ) expressed as percentage of total deposited mass ( DM$_{tot}$ ).  
Different colors correspond to different degrees of aggregation: no aggregation (black), 50\% (yellow), 10\% (blue). B)  i-th mass deposited as aggregate (ADM$_i$) expressed as percentage of i-th deposited mass (DM$_i$ ) .  
Results have been obtained:
i) excluding an area within a distance of about 2 km from the vent
ii) considering a ground loading $> 0.1$ kg/m$^2$ and 
iii) limiting the calculations only to the deposit fallen on the land. } 
\end{figure} 

This is also shown in Fig. \ref{Fig6_aux}A and \ref{Fig6_aux}C, where the TGSD is plotted for different aggregation percentages ($ 0\%$ - black, $ 10\%$ - blue and $ 50\%$ - yellow).   

In Fig. \ref{Fig6_aux}B it is also plotted, for each aggregating i-class (finer than $ \phi$ =3), the mass deposited as aggregate (ADM$_i$) expressed as percentage of the deposited one (DM$_i$).  
As expected the amount of deposited fine particles increases proportionally with aggregation.
Nevertheless, even when $ 10\%$ and $ 50\%$ of aggregation is considered, for   $ \phi=5$ and $ \phi=6$ classes deposit only as aggregates ( ADM$_i$/DM$_i$=100\% ) in the considered domain, 
  
This plot reveals how using these ratio as indicative of the degree of aggregation of sampled particles can be misleading.  
Indeed, the observation of a particular class in the deposit only as aggregate could be erroneously interpreted as evidence that $ 100\%$ of the emitted mass for the i-class aggregated during the transport.

\printbibliography

\end{document}